\documentclass[twocolumn]{aastex631}

\usepackage{color}
\usepackage[normalem]{ulem}

\newcommand{\bjdtdb}{\ensuremath{\rm {BJD_{TDB}}}}

\newcommand{\msun}{\ensuremath{\,M_\Sun}}
\newcommand{\rsun}{\ensuremath{\,R_\Sun}}
\newcommand{\lsun}{\ensuremath{\,L_\Sun}}
\newcommand{\mj}{\ensuremath{\,M_{\rm J}}}
\newcommand{\rj}{\ensuremath{\,R_{\rm J}}}

\newcommand{\fave}{\langle F \rangle}
\newcommand{\fluxcgs}{10$^9$ erg s$^{-1}$ cm$^{-2}$}

\newcommand{\vsini}{\ensuremath{v\sin{i_*}}}

\shorttitle{TOI-4994}
\shortauthors{Rodr\'iguez Mart\'inez et al.}

\begin{document}

\title{Discovery and Characterization of an Eccentric, Warm Saturn Transiting the Solar Analog TOI-4994\footnote{This paper includes data gathered with the 6.5 meter Magellan Telescopes located at Las Campanas Observatory, Chile.}}

\newcommand{\cfa}{Center for Astrophysics \textbar \ Harvard \& Smithsonian, 60 Garden St, Cambridge, MA 02138, USA}
\newcommand{\msu}{Center for Data Intensive and Time Domain Astronomy, Department of Physics and Astronomy, Michigan State University, East
Lansing, MI 48824, USA}
\newcommand{\umich}{Astronomy Department, University of Michigan, 1085 S University Avenue, Ann Arbor, MI 48109, USA}
\newcommand{\utaustin}{Department of Astronomy, The University of Texas at Austin, Austin, TX 78712, USA}
\newcommand{\MIT}{Department of Physics and Kavli Institute for Astrophysics and Space Research, Massachusetts Institute of Technology, Cambridge, MA 02139, USA}
\newcommand{\MITEPS}{Department of Earth, Atmospheric and Planetary Sciences, Massachusetts Institute of Technology,  Cambridge,  MA 02139, USA}
\newcommand{\uflorida}{Department of Astronomy, University of Florida, 211 Bryant Space Science Center, Gainesville, FL, 32611, USA}
\newcommand{\riverside}{Department of Earth and Planetary Sciences, University of California, Riverside, CA 92521, USA}
\newcommand{\usq}{Centre for Astrophysics, University of Southern Queensland, West Street, Toowoomba, QLD 4350, Australia}
\newcommand{\ames}{NASA Ames Research Center, Moffett Field, CA, 94035, USA}
\newcommand{\geneva}{Observatoire astronomique de l\'{}Universit\'{e} de Gen\`{e}ve, 51 Chemin Pegasi, 1290 Versoix, Switzerland}
\newcommand{\uw}{Astronomy Department, University of Washington, Seattle, WA 98195 USA}
\newcommand{\warwick}{Department of Physics, University of Warwick, Gibbet Hill Road, Coventry CV4 7AL, UK}
\newcommand{\warwickceh}{Centre for Exoplanets and Habitability, University of Warwick, Gibbet Hill Road, Coventry CV4 7AL, UK}
\newcommand{\princeton}{Department of Astrophysical Sciences, Princeton University, 4 Ivy Lane, Princeton, NJ, 08544, USA}
\newcommand{\liege}{Space Sciences, Technologies and Astrophysics Research (STAR) Institute, Universit\'e de Li\`ege, 19C All\'ee du 6 Ao\^ut, 4000 Li\`ege, Belgium}
\newcommand{\vanderbilt}{Department of Physics and Astronomy, Vanderbilt University, Nashville, TN 37235, USA}
\newcommand{\fisk}{Department of Physics, Fisk University, 1000 17th Avenue North, Nashville, TN 37208, USA}
\newcommand{\columbia}{Department of Astronomy, Columbia University, 550 West 120th Street, New York, NY 10027, USA}
\newcommand{\toronto}{Dunlap Institute for Astronomy and Astrophysics, University of Toronto, Ontario M5S 3H4, Canada}
\newcommand{\unc}{Department of Physics and Astronomy, University of North Carolina at Chapel Hill, Chapel Hill, NC 27599, USA}
\newcommand{\iac}{Instituto de Astrof\'isica de Canarias (IAC), E-38205 La Laguna, Tenerife, Spain}
\newcommand{\lalaguna}{Departamento de Astrof\'isica, Universidad de La Laguna (ULL), E-38206 La Laguna, Tenerife, Spain}
\newcommand{\louisville}{Department of Physics and Astronomy, University of Louisville, Louisville, KY 40292, USA}
\newcommand{\aavso}{American Association of Variable Star Observers, 49 Bay State Road, Cambridge, MA 02138, USA}
\newcommand{\utokyo}{The University of Tokyo, 7-3-1 Hongo, Bunky\={o}, Tokyo 113-8654, Japan}
\newcommand{\naoj}{National Astronomical Observatory of Japan, 2-21-1 Osawa, Mitaka, Tokyo 181-8588, Japan}
\newcommand{\jstpresto}{JST, PRESTO, 7-3-1 Hongo, Bunkyo-ku, Tokyo 113-0033, Japan}
\newcommand{\astrobiojapan}{Astrobiology Center, 2-21-1 Osawa, Mitaka, Tokyo 181-8588, Japan}
\newcommand{\ctio}{Cerro Tololo Inter-American Observatory, Casilla 603, La Serena, Chile}
\newcommand{\noirlab}{NOIRLab/Cerro Tololo Inter-American Observatory, Casilla 603, La Serena, Chile}
\newcommand{\nexsci}{Caltech IPAC -- NASA Exoplanet Science Institute 1200 E. California Ave, Pasadena, CA 91125, USA}
\newcommand{\ucsc}{Department of Astronomy and Astrophysics, University of
California, Santa Cruz, CA 95064, USA}
\newcommand{\gsfc}{Exoplanets and Stellar Astrophysics Laboratory, Code 667, NASA Goddard Space Flight Center, Greenbelt, MD 20771, USA}
\newcommand{\sgtinc}{SGT, Inc./NASA AMES Research Center, Mailstop 269-3, Bldg T35C, P.O. Box 1, Moffett Field, CA 94035, USA}
\newcommand{\chile}{Center of Astro-Engineering UC, Pontificia Universidad Cat\'olica de Chile, Av. Vicu\~{n}a Mackenna 4860, 7820436 Macul, Santiago, Chile}
\newcommand{\Pontificia}{Facultad de Ingeniería y Ciencias, Universidad Adolfo Ib\'a\~nez, Av. Diagonal las Torres 2640, Pe\~nalol\'en, Santiago, Chile}

\newcommand{\Millennium}{Millennium Institute for Astrophysics, Chile}
\newcommand{\maxplank}{Max-Planck-Institut f\"ur Astronomie, K\"onigstuhl 17, Heidelberg 69117, Germany}
\newcommand{\utdallas}{Department of Physics, The University of Texas at Dallas, 800 West
Campbell Road, Richardson, TX 75080-3021 USA}
\newcommand{\MauryLewin}{Maury Lewin Astronomical Observatory, Glendora, CA 91741, USA}
\newcommand{\umbc}{University of Maryland, Baltimore County, 1000 Hilltop Circle, Baltimore, MD 21250, USA}
\newcommand{\osu}{Department of Astronomy, The Ohio State University, 140 West 18th Avenue, Columbus, OH 43210, USA}
\newcommand{\MITAA}{Department of Aeronautics and Astronautics, MIT, 77 Massachusetts Avenue, Cambridge, MA 02139, USA}
\newcommand{\openu}{School of Physical Sciences, The Open University, Milton Keynes MK7 6AA, UK}
\newcommand{\swarthmore}{Department of Physics and Astronomy, Swarthmore College, Swarthmore, PA 19081, USA}
\newcommand{\seti}{SETI Institute, Mountain View, CA 94043, USA}
\newcommand{\lehigh}{Department of Physics, Lehigh University, 16 Memorial Drive East, Bethlehem, PA 18015, USA}
\newcommand{\utah}{Department of Physics and Astronomy, University of Utah, 115 South 1400 East, Salt Lake City, UT 84112, USA}
\newcommand{\USNA}{Department of Physics, United States Naval Academy, 572C Holloway Rd., Annapolis, MD 21402, USA}
\newcommand{\DTM}{Department of Terrestrial Magnetism, Carnegie Institution for Science, 5241 Broad Branch Road, NW, Washington, DC 20015, USA}
\newcommand{\UPenn}{The University of Pennsylvania, Department of Physics and Astronomy, Philadelphia, PA, 19104, USA}
\newcommand{\montana}{Department of Physics and Astronomy, University of Montana, 32 Campus Drive, No. 1080, Missoula, MT 59812 USA}
\newcommand{\psu}{Department of Astronomy \& Astrophysics, The Pennsylvania State University, 525 Davey Lab, University Park, PA 16802, USA}
\newcommand{\psust}{Center for Exoplanets and Habitable Worlds, The Pennsylvania State University, 525 Davey Lab, University Park, PA 16802, USA}
\newcommand{\Kutztown}{Department of Physical Sciences, Kutztown University, Kutztown, PA 19530, USA}
\newcommand{\udel}{Department of Physics \& Astronomy, University of Delaware, Newark, DE 19716, USA}
\newcommand{\Westminster}{Department of Physics, Westminster College, New Wilmington, PA 16172}
\newcommand{\steward}{Department of Astronomy and Steward Observatory, University of Arizona, Tucson, AZ 85721, USA}
\newcommand{\saao}{South African Astronomical Observatory, PO Box 9, Observatory, 7935, Cape Town, South Africa}
\newcommand{\salt}{Southern African Large Telescope, PO Box 9, Observatory, 7935, Cape Town, South Africa}
\newcommand{\ssl}{Societ\`{a} Astronomica Lunae, Italy}
\newcommand{\spot}{Spot Observatory, Nashville, TN 37206, USA}
\newcommand{\txamGP}{George P.\ and Cynthia Woods Mitchell Institute for Fundamental Physics and Astronomy, Texas A\&M University, College Station, TX77843 USA}
\newcommand{\txam}{Department of Physics and Astronomy, Texas A\&M university, College Station, TX 77843 USA}
\newcommand{\wellesley}{Department of Astronomy, Wellesley College, Wellesley, MA 02481, USA}
\newcommand{\Wesleyan}{Astronomy Department and Van Vleck Observatory, Wesleyan University, Middletown, CT 06459, USA}
\newcommand{\inaf}{INAF -- Osservatorio Astronomico di Padova, Vicolo dell'Osservatorio 5, I-35122, Padova, Italy}
\newcommand{\byu}{Department of Physics and Astronomy, Brigham Young University, Provo, UT 84602, USA}
\newcommand{\Hazelwood}{Hazelwood Observatory, Churchill, Victoria, Australia}
\newcommand{\austinstate}{Department of Physics, Engineering and Astronomy, Stephen F. Austin State University, 1936 North St, Nacogdoches, TX 75962, USA}
\newcommand{\pest}{Perth Exoplanet Survey Telescope}
\newcommand{\Winer}{Winer Observatory, PO Box 797, Sonoita, AZ 85637, USA}
\newcommand{\icpo}{Ivan Curtis Private Observatory}
\newcommand{\elsauce}{El Sauce Observatory, Chile}
\newcommand{\crow}{Atalaia Group \& CROW Observatory, Portalegre, Portugal}
\newcommand{\dfus}{Dipartimento di Fisica ``E.R.Caianiello'', Universit\`a di Salerno, Via Giovanni Paolo II 132, Fisciano 84084, Italy}
\newcommand{\indfn}{Istituto Nazionale di Fisica Nucleare, Napoli, Italy}
\newcommand{\sotes}{Gabriel Murawski Private Observatory (SOTES)}
\newcommand{\lco}{Las Cumbres Observatory Global Telescope, 6740 Cortona Dr., Suite 102, Goleta, CA 93111, USA}
\newcommand{\ucsb}{Department of Physics, University of California, Santa Barbara, CA 93106-9530, USA}
\newcommand{\yale}{Department of Astronomy, Yale University, 52 Hillhouse Avenue, New Haven, CT 06511, USA}
\newcommand{\eso}{European Southern Observatory, Alonso de C\'ordova 3107, Vitacura, Casilla 19001, Santiago, Chile}
\newcommand{\stsci}{Space Telescope Science Institute, Baltimore, MD 21218, USA}
\newcommand{\keele}{Astrophysics Group, Keele University, Staffordshire ST5 5BG, UK}
\newcommand{\gsfcsellers}{GSFC Sellers Exoplanet Environments Collaboration, NASA Goddard Space Flight Center, Greenbelt, MD 20771 }
\newcommand{\usno}{U.S. Naval Observatory, Washington, DC 20392, USA}
\newcommand{\kansas}{Department of Physics and Astronomy, University of Kansas, 1251 Wescoe Hall Dr., Lawrence, KS 66045, USA}
\newcommand{\gmu}{George Mason University, 4400 University Drive MS 3F3, Fairfax, VA 22030, USA}
\newcommand{\unsw}{Exoplanetary Science at UNSW, School of Physics, UNSW Sydney, NSW 2052, Australia}
\newcommand{\sifa}{School of Physics, Sydney Institute for Astronomy (SIfA), The University of Sydney, NSW 2006, Australia}
\newcommand{\nanjing}{School of Astronomy and Space Science, Key Laboratory of Modern Astronomy and Astrophysics in Ministry of Education, Nanjing University, Nanjing 210046, Jiangsu, China}
\newcommand{\berkely}{Department of Astronomy, University of California Berkeley, Berkeley, CA 94720-3411, USA}
\newcommand{\bhicfa}{Black Hole Initiative at Harvard University, 20 Garden Street, Cambridge, MA 02138, USA}
\newcommand{\Silesian}{Department of Electronics, Electronical Engineering and Microelectronics, Silesian University of Techhnology Akademicka 16, 44-100 Gliwice, Poland}
\newcommand{\Patashnick}{Patashnick Voorheesville Observatory, Voorheesville, NY 12186, USA}
\newcommand{\austincollege}{Physics Department, Austin College, 900 North Grand Avenue, Sherman TX 75090, USA}
\newcommand{\Tsinghua}{Department of Astronomy, Tsinghua University, Beijing 100084, China}
\newcommand{\Tsinghuaschool}{Tsinghua International School, Beijing 100084, China}
\newcommand{\chinaNAO}{National Astronomical Observatories, Chinese Academy of Sciences, 20A Datun Road, Chaoyang District, Beijing 100012, China}
\newcommand{\Tautenburg}{Th{\"u}ringer Landessternwarte Tautenburg, Sternwarte 5, 07778 Tautenburg, Germany}
\newcommand{\brierfield}{Brierfield Observatory, New South Wales, Australia}
\newcommand{\Indiana}{Indiana University Department of Astronomy, SW319, 727 E 3rd Street, Bloomington, IN 47405 USA}
\newcommand{\wisconsin}{Department of Astronomy, University of Wisconsin-Madison, Madison, WI 53706, USA}
\newcommand{\protologic}{Proto-Logic Consulting LLC, Washington, DC 20009, USA}
\newcommand{\ASTRAVEO}{ASTRAVEO LLC, PO Box 1668, MA 01931}
\newcommand{\TJHS}{Thomas Jefferson High School, 6560 Braddock Rd, Alexandria, VA 22312 USA}
\newcommand{\ucatchile}{Instituto de Astrof\'isica, Facultad de F\'isica, Pontificia Universidad Cat\'olica de Chile}
\newcommand{\lasa}{Liberal Arts and Science Academy, Austin, Texas 78724, USA}
\newcommand{\gemini}{Gemini Observatory/NSF’s NOIRLab, 670 N. A’ohoku Place, Hilo, HI, 96720, USA}
\newcommand{\umd}{Department of Astronomy, University of Maryland, College Park, College Park, MD}
\newcommand{\Bern}{Physikalisches Institut, University of Bern, Gesellschaftsstrasse 6, 3012 Bern, Switzerland}

\newcommand{\eberly}{\altaffiliation{Eberly Research Fellow}}
\newcommand{\torres}{\altaffiliation{Juan Carlos Torres Fellow}}
\newcommand{\sagan}{\altaffiliation{NASA Sagan Fellow}}
\newcommand{\bernoulli}{\altaffiliation{Bernoulli fellow}}
\newcommand{\gruber}{\altaffiliation{Gruber fellow}}
\newcommand{\kavli}{\altaffiliation{Kavli Fellow}}
\newcommand{\peg}{\altaffiliation{51 Pegasi b Fellow}}
\newcommand{\pappalardo}{\altaffiliation{Pappalardo Fellow}}
\newcommand{\hubble}{\altaffiliation{NASA Hubble Fellow}}
\newcommand{\nsf}{\altaffiliation{National Science Foundation Graduate Research Fellow}}

\newcommand{\maxplanck}{Max-Planck-Institut für Astronomie,Königstuhl 17, D-69117 Heidelberg, Germany}

\newcommand{\Sofia}{Department of Astronomy, Sofia University ``St Kliment Ohridski'', 5 James Bourchier Blvd, BG-1164 Sofia, Bulgaria}
 
\newcommand{\Heidelberg}{Landessternwarte, Zentrum f\"ur Astronomie der Universit\"at Heidelberg, K\"onigstuhl 12, D-69117 Heidelberg, Germany}

\newcommand{\Adolfo}{Facultad de Ingeniera y Ciencias, Universidad Adolfo Ib\'{a}\~{n}ez, Av. Diagonal las Torres 2640, Pe\~{n}alol\'{e}n, Santiago, Chile}
\newcommand{\Millenium}{Millenium Institute of Astrophysics, Santiago, Chile}

\correspondingauthor{Romy Rodr\'iguez} 
\email{romy.rodriguez@cfa.harvard.edu}

\author[0000-0003-1445-9923]{Romy Rodr\'iguez Mart\'inez} 
\affiliation{\cfa}

\author[0000-0003-3773-5142]{Jason D. Eastman}
\affiliation{\cfa}

\author[0000-0001-6588-9574]{Karen A.\ Collins}
\affiliation{\cfa}

\author[0000-0001-8812-0565]{Joseph E. Rodriguez}
\affiliation{\msu}

\author[0000-0002-9003-484X]{David Charbonneau}
\affiliation{\cfa}

\author[0000-0002-8964-8377]{Samuel N. Quinn}
\affiliation{\cfa}

\author[0000-0001-9911-7388]{David W. Latham} 
\affiliation{\cfa}

\author[0000-0002-0619-7639]{Carl Ziegler}
\affiliation{\austinstate}

\author[0000-0002-9158-7315]{Rafael Brahm}
\affiliation{\Adolfo}
\affiliation{\Millennium}

\author[0000-0002-0692-7822]{Tyler R. Fairnington}
\affiliation{\usq}

\author[0000-0003-2417-7006]{Solène Ulmer-Moll}
\affiliation{\geneva}
\affiliation{\Bern}

\author[0000-0002-3481-9052]{Keivan G.\ Stassun}
\affiliation{Department of Physics and Astronomy, Vanderbilt University, Nashville, TN 37235, USA}

\author[0000-0002-3503-3617]{Olga Suarez}
\affiliation{Universit\'e C\^ote d'Azur, Observatoire de la C\^ote d'Azur, CNRS, Laboratoire Lagrange, Bd de l'Observatoire, CS 34229, 06304 Nice cedex 4, France}

\author[0000-0002-7188-8428]{Tristan Guillot}
\affiliation{Universit\'e C\^ote d'Azur, Observatoire de la C\^ote d'Azur, CNRS, Laboratoire Lagrange, Bd de l'Observatoire, CS 34229, 06304 Nice cedex 4, France}

\author[0000-0002-5945-7975]{Melissa J.\ Hobson}
\affiliation{\geneva}

\author[0000-0002-4265-047X]{Joshua N.\ Winn}
\affiliation{Department of Astrophysical Sciences, Princeton University, Princeton, NJ 08544, USA}

\author[0000-0001-8401-4300]{Shubham Kanodia}
\affiliation{Carnegie Science Earth and Planets Laboratory, 5241 Broad Branch Road, NW, Washington, DC 20015, USA}

\author[0000-0001-8355-2107]{Martin Schlecker}
\affiliation{\steward}


\author[0000-0003-1305-3761]{R.P. Butler}
\affiliation{Carnegie Science Earth and Planets Laboratory, 5241 Broad Branch Road, NW, Washington, DC 20015, USA}

\author[0000-0002-5226-787X]{Jeffrey D. Crane}
\affiliation{Carnegie Science Earth and Planets Laboratory, 5241 Broad Branch Road, NW, Washington, DC 20015, USA}

\author[0000-0002-8681-6136]{Steve Shectman}
\affiliation{Carnegie Science Earth and Planets Laboratory, 5241 Broad Branch Road, NW, Washington, DC 20015, USA}

\author[0009-0008-2801-5040]{Johanna K. Teske}
\affiliation{Carnegie Science Earth and Planets Laboratory, 5241 Broad Branch Road, NW, Washington, DC 20015, USA}

\author[0000-0003-0412-9664]{David Osip}
\affiliation{Carnegie Science Earth and Planets Laboratory, 5241 Broad Branch Road, NW, Washington, DC 20015, USA}

\author{Yuri Beletsky}
\affiliation{Carnegie Science Earth and Planets Laboratory, 5241 Broad Branch Road, NW, Washington, DC 20015, USA}

\author[0000-0002-1357-9774]{Matthew P. Battley}
\affiliation{\geneva}

\author[0000-0002-4797-2419]{Angelica Psaridi}
\affiliation{\geneva}

\author[0000-0001-8504-283X]{Pedro Figueira}
\affiliation{\geneva}

\author[0000-0001-9699-1459]{Monika Lendl}
\affiliation{\geneva}

\author{François Bouchy }
\affiliation{\geneva}

\author{St\'ephane Udry}
\affiliation{\geneva}

\author[0000-0001-9269-8060]{Michelle Kunimoto}
\affiliation{Department of Physics and Kavli Institute for Astrophysics and Space Research, Massachusetts Institute of Technology, 77 Massachusetts Avenue, Cambridge, MA 02139, USA}


\author{Djamel Mékarnia}
\affiliation{Universit\'e C\^ote d'Azur, Observatoire de la C\^ote d'Azur, CNRS, Laboratoire Lagrange, Bd de l'Observatoire, CS 34229, 06304 Nice cedex 4, France}

\author{Lyu Abe}
\affiliation{Universit\'e C\^ote d'Azur, Observatoire de la C\^ote d'Azur, CNRS, Laboratoire Lagrange, Bd de l'Observatoire, CS 34229, 06304 Nice cedex 4, France}


\author[0000-0002-0236-775X]{Trifon Trifonov}
\affiliation{\maxplanck}
\affiliation{\Sofia}
\affiliation{\Heidelberg}

\author{Marcelo Tala Pinto}
\affiliation{\Adolfo}
\affiliation{\Millennium}

\author[0000-0003-3130-2768]{Jan Eberhardt}
\affiliation{\maxplanck}

\author[0000-0001-9513-1449]{Nestor Espinoza}
\affiliation{\stsci}

\author[0000-0002-1493-300X]{Thomas Henning}
\affiliation{\Heidelberg}

\author[0000-0002-5389-3944]{Andr\'es Jord\'an}
\affiliation{\Adolfo}

\author[0000-0003-3047-6272]{Felipe I.\ Rojas}
\affiliation{Instituto de Astrof\'isica, Facultad de F\'isica, Pontificia Universidad Cat\'olica de Chile, Av. Vicu\~{n}a Mackenna 4860, Santiago, Chile.}

\author[0000-0003-1464-9276]{Khalid Barkaoui}
\affiliation{Astrobiology Research Unit, Universit\'e de Li\`ege, 19C All\'ee du 6 Ao\^ut, 4000 Li\`ege, Belgium}
\affiliation{Department of Earth, Atmospheric and Planetary Science, Massachusetts Institute of Technology, 77 Massachusetts Avenue, Cambridge, MA 02139, USA}
\affiliation{Instituto de Astrof\'isica de Canarias (IAC), Calle V\'ia L\'actea s/n, 38200, La Laguna, Tenerife, Spain}

\author[0009-0009-5132-9520]{Howard M. Relles}
\affiliation{\cfa}

\author{Gregor Srdoc}
\affil{Kotizarovci Observatory, Sarsoni 90, 51216 Viskovo, Croatia}

\author[0000-0003-2781-3207]{Kevin I.\ Collins}
\affiliation{George Mason University, 4400 University Drive, Fairfax, VA, 22030 USA}

\author[0000-0002-6892-6948]{Sara Seager}
\affiliation{Department of Physics and Kavli Institute for Astrophysics and Space Research, Massachusetts Institute of Technology, Cambridge, MA 02139, USA}
\affiliation{Department of Earth, Atmospheric and Planetary Sciences, Massachusetts Institute of Technology, Cambridge, MA 02139, USA}
\affiliation{Department of Aeronautics and Astronautics, MIT, 77 Massachusetts Avenue, Cambridge, MA 02139, USA}

\author[0000-0002-1836-3120]{Avi Shporer}
\affiliation{Department of Physics and Kavli Institute for Astrophysics and Space Research, Massachusetts Institute of Technology, 77 Massachusetts Avenue, Cambridge, MA 02139, USA}

\author{Michael~Vezie}
\affiliation{Department of Physics and Kavli Institute for Astrophysics and Space Research, Massachusetts Institute of Technology, Cambridge, MA 02139, USA}

\author{Christina Hedges}

\affiliation{NASA Goddard Space Flight Center, 8800 Greenbelt Rd, Greenbelt, MD 20771, USA}

\author[0000-0002-4510-2268]{Ismael~Mireles}
\affiliation{Department of Physics and Astronomy, University of New Mexico, 210 Yale Blvd NE, Albuquerque, NM 87106, USA}

\begin{abstract}

We present the detection and characterization of TOI-4994b (TIC 277128619b), a warm Saturn-sized planet discovered by the NASA Transiting Exoplanet Survey Satellite (TESS). TOI-4994b transits a G-type star (V = 12.6 mag) with a mass, radius, and effective temperature of $M_{\star} =1.005^{+0.064}_{-0.061} M_{\odot}$, 
$R_{\star} = 1.055^{+0.040}_{-0.037} R_{\odot}$, and $T_{\rm eff} = 5640 \pm 110$ K. We obtained follow-up ground-based photometry from the Las Cumbres Observatory (LCO) and the Antarctic Search for Transiting ExoPlanets (ASTEP) telescopes, and we confirmed the planetary nature of TOI-4994b with multiple radial velocity observations from the PFS, CHIRON, HARPS, FEROS, and CORALIE instruments. From a global fit to the photometry and radial velocities, we determine that TOI-4994b is in a 21.5-day, eccentric orbit ($e = 0.32 \pm 0.04$) and has a mass of $M_{P}= 0.280^{+0.037}_{-0.034} M_{J}$, a radius of $R_{P}= 0.762^{+0.030}_{-0.027}R_{J}$, and a Saturn-like bulk density of $\rho_{p} = 0.78^{+0.16}_{-0.14}$ $\rm g/cm^3$. We find that TOI-4994 is a potentially viable candidate for follow-up stellar obliquity measurements. TOI-4994b joins the small sample of warm Saturn analogs and thus sheds light on our understanding of these rare and unique worlds.  


\end{abstract}

\keywords{exoplanets - techniques: radial velocities - techniques: spectroscopic - techniques:
photometric - methods: observational}

\section{Introduction} \label{sec:intro}

The number of known giant exoplanets has vastly expanded since the discovery of the first transiting giant planet \citep{charbonneau:2000,henry:2000} with the advent of many wide-field transit and radial velocity (RV) surveys (e.g., \citealt{Bakos:2007,bakos:2013,pollacco:2006,Pepper:2007, Borucki:2010}). However, despite the thousands of giant planet discoveries brought about by these dedicated surveys, there are still many open questions regarding the formation mechanisms, composition, and evolution of these planets. One such longstanding problem in planetary astrophysics is whether close-in giant planets form in situ or instead form at larger orbital separations beyond the snowline and subsequently migrate inwards through interactions with the protoplanetary disk (e.g., \citealt{Kley:2012,Nelson:2018}). In addition, the migration pathways of these planets are poorly understood (see, e.g., \cite{Winn:2015}, \cite{Dawson:2018} and \citealt{Schulte:2024}). 

Transiting giant planets offer important clues to some of these problems, as they are comparatively easier to detect than small planets and are rich targets for characterization. First, they enable precise measurements of masses, radii, and orbital properties, providing valuable insights into their bulk composition and interior structure \citep{Fortney:2007,Fortney:2010}. The study of these objects refines and contextualizes our knowledge of the giant planets in our Solar System, while allowing us to explore the extreme physics of other types of planets that are not present within it. The relatively larger scale heights of giant planets (i.e., planet-to-star radius ratio) make them excellent targets for follow-up atmospheric characterization, providing clues about their heavy-element mass, interior structure and formation histories \citep{Madhusudhan:2014, Thorngren:2016}. These planets are also particularly well-suited for measurements of the sky-projected spin-orbit angle with the Rossiter-McLaughlin effect or Doppler Tomography \citep{Gaudi:2007}. 

Warm giant planets, which have equilibrium temperatures below $\sim$1000 K, or orbital periods between 10--200 d ($a$ $\gtrsim$ 0.1 AU), are uniquely interesting because, at their larger distances from their host stars, they are likely not subjected to the same physical processes that inflate the atmospheres of hot Jupiters \citep{Fortney:2010,Jordan:2019,Muller:2023}. Therefore, their observed properties are likely primordial and enable comparisons between planets at close distances from their host stars. However, the number of well-characterized warm transiting giants is still small, partly because the transit and RV methods are both biased towards planets on shorter periods. Additionally, most of the warm giants discovered from space missions like $Kepler$ and CoRoT orbit fainter stars that are not especially amenable to follow-up observations. In fact, as of August 9, 2024, warm giants comprise only 17\% (180 planets) of all known giants.\footnote{Data from the NASA Exoplanet Archive.} 

The NASA Transiting Exoplanet Survey Satellite (TESS; \citealt{Ricker:2015}) has been tremendously successful at finding all kinds of exoplanets, and it is expected to find over 250 giant planets with 2-minute cadence \citep{Barclay:2018}.  In this paper, we present the discovery of the warm Saturn TOI-4994b, which orbits a moderately bright (V = 12.6 mag) G-type star on an eccentric, 21.5-day period orbit, discovered with data from TESS. The signals were followed up and confirmed as planetary with a suite of ground-based photometry from Las Cumbres Observatory \citep{Brown:2013} and the ASTEP-Antarctica observatory \citep{Daban:2010}, and radial velocity observations from multiple spectrographs. The unique properties of Saturn analogs along with their relative rarity make them valuable laboratories to test and refine theories of planet formation. This paper is structured as follows. In Section~\ref{sec:observations}, we provide confirmation of the planetary nature of TOI-4994b from our photometric and spectroscopic observations. In Section~\ref{sec:analysis}, we present our analysis of the data. In Section~\ref{sec:results_discussion}, we present our results and the properties of the planet in the context of the known exoplanet population; and we finally conclude in Section~\ref{sec:conclusions}. 

\begin{table}
\scriptsize
\setlength{\tabcolsep}{2pt}
\centering
\caption{Stellar Properties from the Literature.}
\label{tbl:LitProps}
\begin{tabular}{lcc}
\hline
 \hline
\textbf{Parameter} & \textbf{Value} & \textbf{Source}\\
\hline 

\smallskip\\\multicolumn{2}{l}{\textbf{Identifiers}}&\smallskip\\

TIC ID & 	277128619 & 1	\\
TOI & 	4994 & 1	\\
2MASS & J18523181-7826041 & 1	\\
Gaia DR3 & 6364463103935391232 & 1 \\

\smallskip\\\multicolumn{2}{l}{\textbf{Coordinates and Proper Motion}}&\smallskip\\

Right Ascension (RA) & 	18:52:31.81 & 2	\\
Declination (Dec)& -78:26:03.95 & 2	\\
$\mu_{\alpha}$ (mas yr$^-1$) & $0.897 \pm 0.009$ & 2 \\
$\mu_{\delta}$ (mas yr$^-1$) & $9.90 \pm 0.01$ & 2 \\

Gaia Parallax (mas) & $3.022^{+0.022}_{-0.023}$ &  2 \\
Distance (pc) & $330.9^{+2.5}_{-2.4}$
 & 3 \\

\smallskip\\\multicolumn{2}{l}{\textbf{Broadband Magnitudes}}&\smallskip\\

Gaia $G$ mag.     &12.4792 $\pm$0.0001& 2\\
Gaia ${\rm G_{BP}}$ mag.  &12.913 $\pm$ 0.001& 2\\
Gaia ${\rm G_{RP}}$ mag. &11.9044 $\pm$ 0.0007& 2\\
TESS mag.    & 11.954 $\pm$ 0.006 & 4
\\
\\

2MASS J mag. & 11.221 $\pm$ 0.021 & 5	\\
2MASS H mag. & 10.85 $\pm$ 0.02   &  5	\\
2MASS ${\rm K_S}$ mag. & 10.797 $\pm$ 0.022 &  5	\\
\\
\textit{WISE 1} mag. & 10.734 $\pm$ 0.023 & 6	\\
 \textit{WISE 2} mag. & 10.793 $\pm$ 0.021 &  6 \\
\textit{WISE 3} mag. &  10.684 $\pm$ 0.068 & 6	\\


 \\
\hline
\end{tabular}
\begin{flushleft}
 \footnotesize{ \textbf{\textsc{NOTES:}}
References: $^1$SIMBAD Astronomical Database, $^2$\citet{Gaia:2023}, $^3$This work, $^4$TIC-v8.2 \citep{stassun:2019}, 
$^5$\citet{Cutri:2003}, $^6$\citet{Zacharias:2017}. \\
}
\end{flushleft}
\end{table}

\section{Observations} \label{sec:observations}

TOI-4994 (TIC 277128619) is a moderately bright (V=12.6 mag) G-type star located at RA = 18$^{h}$52$^{m}$31$\fs$81 and Dec = $-$78$\degr$26$\arcmin$03$\arcsec$95, at a distance of 330.9 $\pm$ 2.45 pc (see Table~\ref{tbl:LitProps} and Section~\ref{sec:analysis} for more on the properties of the host). A transit-like signal with a period of 21.5 days was first detected in Sector 12 of TESS and subsequently confirmed with follow-up photometry from the ground. We describe these and the radial velocity observations in detail in the following subsections.

\subsection{\rm \textbf{TESS Photometry}}
\label{sec:TESS_photometry}

Since its launch in 2018, the TESS mission has found thousands of transiting exoplanet candidates around the nearest and brightest dwarf stars. TESS has four cameras that survey 95\% of the sky. Currently, over 400 exoplanets have been confirmed or statistically validated, and we expect many more to follow. 

TOI-4994b (TIC 277128619b) was first observed in Sectors 12 and 13 in June-July 2019 by Camera 3 at 30-minute cadence, and then again in Sectors 27 (July 2020) and 39 (June 2021) at 10-minute cadence, and finally in Sectors 66 and 67 in June-July 2023 at 2-minute cadence \citep{Stumpe2012,Stumpe2014,Smith}. A total of 11 TESS transits were observed across these sectors. The target will also be re-observed in 2025 in June and July. Data from all six sectors were used in our analysis. 

\begin{figure*}
\begin{center}
    \includegraphics[width=1.0\textwidth]{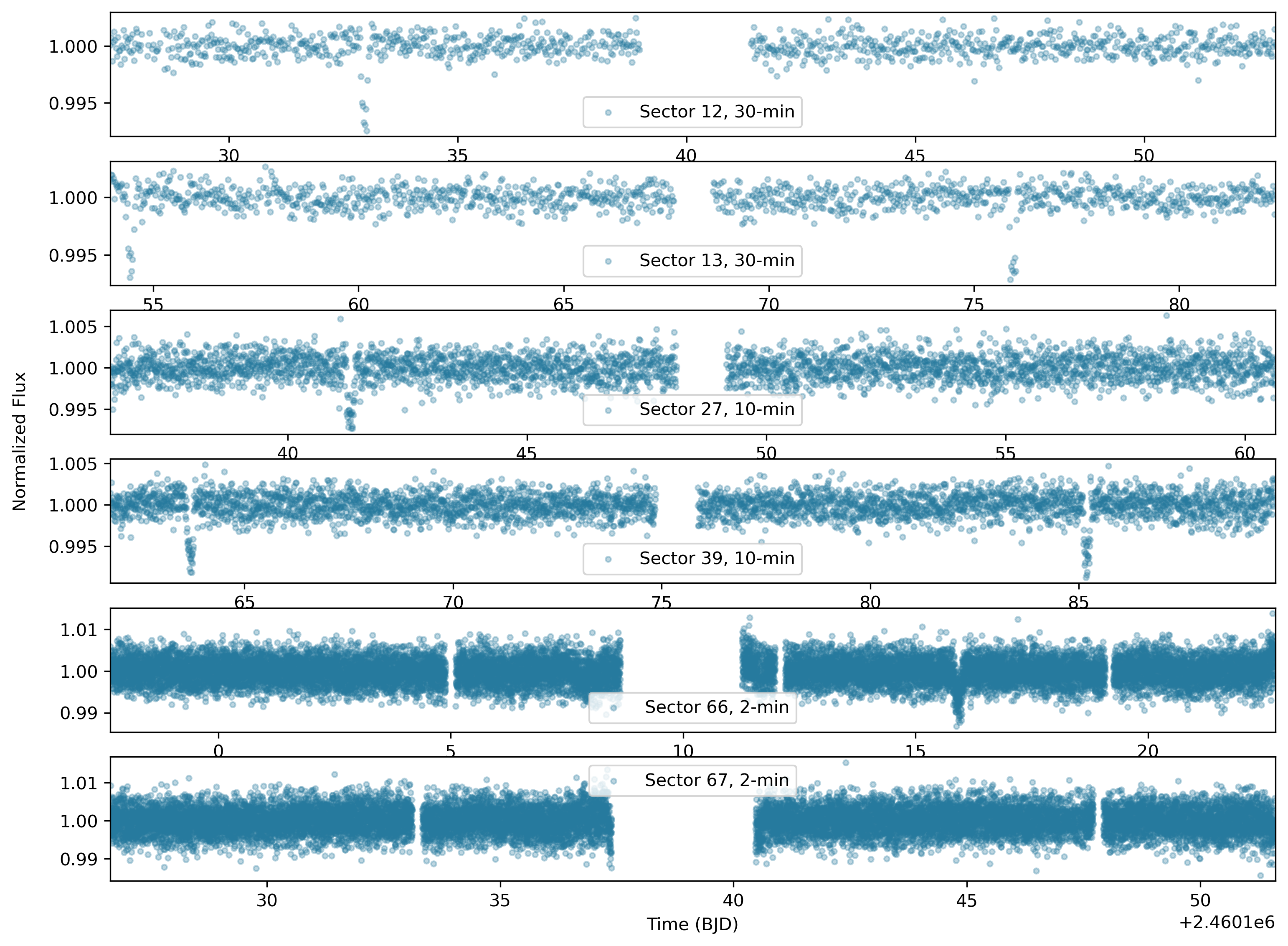}
\end{center}
\caption{Time-series photometry of all 6 TESS sectors used in our analysis.}
\label{fig:timeseries_tess}
\end{figure*}

\begin{figure}[!ht]
\vspace{.1in}
\centering
\includegraphics[width=1\linewidth]{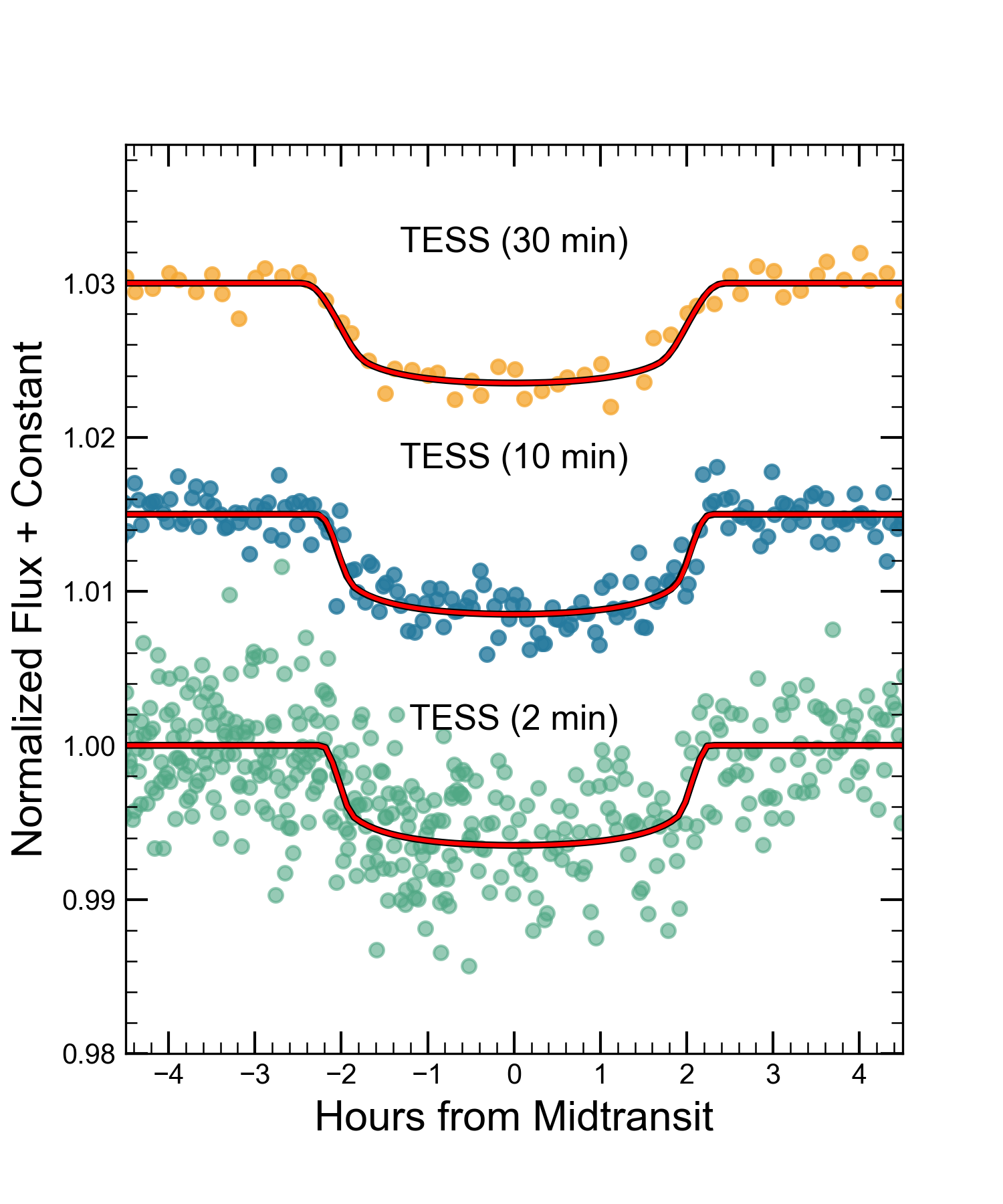}

\caption{TESS light curves of TOI-4994 at 30 (top), 10 (middle), and 2 (bottom) minute cadences, combining 6 different sectors. The red lines show the best-fit {\tt EXOFASTv2} models, phase-folded on the orbital period.}
\label{fig:TESS_lcs} 
\end{figure}

The TESS data were processed and reduced with the NASA Science Processing Operations Center pipeline (SPOC; \citealt{Jenkins:2016,Caldwell2020}), which searches for and identifies transit-like signals in the data. A threshold-crossing event was detected at 21.5 days using a Box Least Squares (BLS) algorithm, identifying the target as a TESS Object of Interest (TOI). The light curves were then downloaded using the {\tt lightkurve} package \citep{lightkurve}, which retrieves data from the Barbara A. Mikulski Archive for Space Telescopes (MAST\footnote{https://mast.stsci.edu/portal/Mashup/Clients/Mast/Portal.html}). Finally, the light curves were flattened using {\tt keplersplinev2}\footnote{https://github.com/avanderburg/keplersplinev2} \citep{Vanderburg:2014}, which removes any out-of-transit variability due to stellar activity. The timeseries and phase-folded TESS transit light curves are shown in Figures~\ref{fig:timeseries_tess} and~\ref{fig:TESS_lcs}.

\subsection{\rm \textbf{Ground-based Photometry}}\label{ground_based}

The TESS pixel scale is $\sim $21$\arcsec$ pixel$^{-1}$ and photometric apertures typically extend out to roughly 1 arcminute, generally causing multiple stars to blend in the photometric aperture. To determine the true source of the TESS detection, we acquired ground-based time series follow-up photometry of the field around TOI-4994 as part of the TESS Follow-up Observing Program \citep[TFOP;][]{collins:2019}\footnote{https://tess.mit.edu/followup}. We used the {\tt TESS Transit Finder}, which is a customized version of the {\tt Tapir} software package \citep{Jensen:2013}, to schedule our transit observations. 

\subsubsection{LCOGT}

\begin{figure}[!ht]
\vspace{.1in}
\centering
\includegraphics[width=1\linewidth]{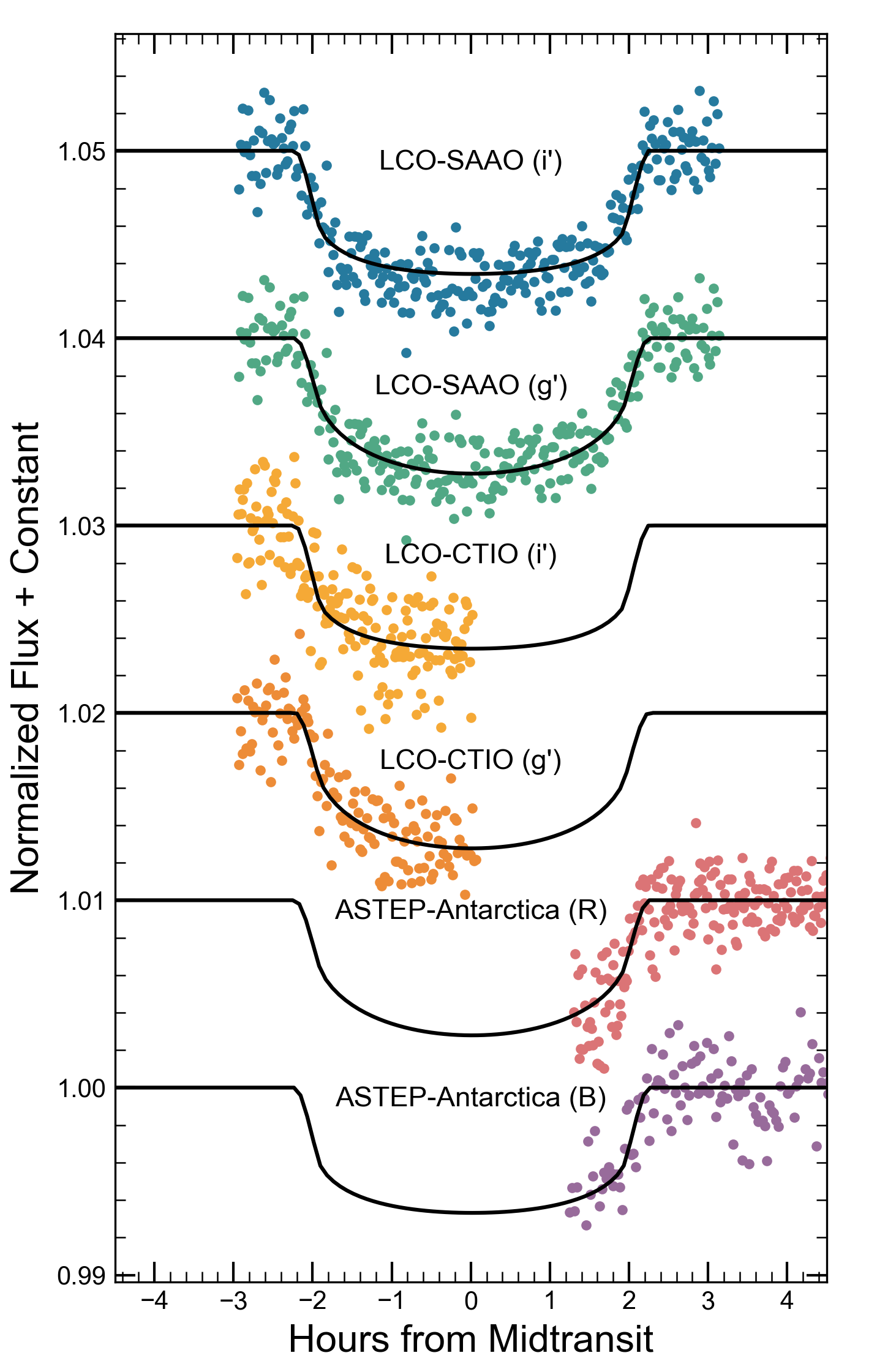}

\caption{Light curves of TOI-4994 from the LCO and ASTEP-Antarctica ground-based observatories. }
\label{fig:Ground_based} 
\end{figure}

Follow-up light curve observations were taken with the Las Cumbres Observatory Global Telescope \citep[LCOGT;][]{Brown:2013} 1-m network nodes at the South African Astronomical Observatory (SAAO) in Cape Town, South Africa and Cerro Tololo Interamerican Observatory (CTIO) in Chile. A full transit was observed simultaneously in $g^{\prime}$ and $i^{\prime}$ bands on July 11, 2023 from SAAO and an ingress observation was conducted simultaneously in $g^{\prime}$ and $i^{\prime}$ bands on May 8, 2023 from CTIO. The 1-m telescopes are equipped with $4096\times4096$ SINISTRO cameras with an image scale of $0\farcs389$ per pixel, resulting in a field of view of $26\arcmin\times26\arcmin$. The images were calibrated by the standard LCOGT {\tt BANZAI} pipeline \citep{McCully:2018}, and differential photometric data were extracted using {\tt AstroImageJ} \citep{Collins:2017}. Circular photometric apertures with a $4\farcs3$ radius were used, effectively excluding most of the flux from the nearest known neighbor in the \textit{Gaia} DR3 catalog, which is 6$\arcsec$ west of TOI-4994 and is 6.6 magnitudes fainter in the TESS band. The light curve data are included in the global modeling described in Section~\ref{sec:analysis} and the phase-folded transits are shown in Figure~\ref{fig:Ground_based}.

\subsubsection{ASTEP-ANTARCTICA}

We obtained follow-up photometric observations of TOI-4994 with the Antarctic Search for Transiting ExoPlanets (ASTEP) telescope. ASTEP is a 0.4-m telescope located at Concordia station in Antarctica \citep{Guillot:2015,Mekarnia:2016}. It is equipped with two back-illuminated cameras operating in the $B$+$V$ bands, similar to Gaia-$B$ (FLI Kepler KL400 sCMOS camera, $2048\times2048$ pixels), and in a red band close to the $Gaia-R$ band (Andor iKon-L 936 CCD camera, $2048\times2048$ pixels). These cameras have an image scale of 1.05\,\arcsec\ and 1.30\,\arcsec\,pixel$^{-1}$ respectively, resulting in corrected fields of view of $36\times36$ arcmin$^2$ and $44\times44$ arcmin$^2$ \citep{schmider:2022}. The data are processed on-site using IDL \citep{Mekarnia:2016} and Python aperture photometry pipelines \citep{Dransfield:2022}. 

Four observations of TOI-4994 were carried out with ASTEP, all under good weather conditions.  
A full transit of TOI-4994b was observed on UT 2022 August 23 in the $R$-band and on UT 2023 May 8 in the $B$ and $R$ bands, while two egress observations were observed on UT 2023 March 26 and on UT 2023 May 29, both in the $B$ and $R$ bands.The phase-folded transits are shown in Figure~\ref{fig:Ground_based}.

\subsection{\rm \textbf{Spectroscopic Observations}}
\label{sec:spectroscopy}

We obtained follow-up radial velocity observations from multiple spectrographs to constrain the mass and orbit of TOI-4994b, which we describe in detail below. To assess whether the star is magnetically active, we calculated its chromospheric activity index, $R'_{\rm HK}$, which is based on the flux emitted by $\rm Ca_{\rm II}$  H and K lines. The $R'_{\rm HK}$ index is the fraction of the star's bolometric luminosity emitted in these lines through magnetic activity. Using the empirical NUV-$R'_{\rm HK}$ relations from \citet{Findeisen:2010}, we obtained log $R'_{\rm HK}$ $-5.06 \pm 0.05$, which is near the mininum value of the chromospheric activity index for stars with solar metallicity of log $R'_{\rm HK} = -5.08$. 
We therefore treat TOI-4994 as an inactive star. All the RV datasets are shown as a function of time and phase-folded on the best-fit period in Figures~\ref{fig:RVs} and~\ref{fig:PFS_HARPS} and listed in the Appendix (\ref{sec:appendix}).

\subsubsection{Planet Finder Spectrograph (PFS)}

We collected 15 RV measurements of TOI-4994 between UT 2023 July 28 and UT 2023 September 25 with the Carnegie \textit{Planet Finder Spectrograph} (PFS;  \citealt{Crane:2006,Crane:2008,Crane:2010}). PFS is a high-precision echelle spectrograph attached to the 6.5-m Magellan Clay telescope at Las Campanas Observatory in Chile. It has a spectral resolution of 130,000 and covers the 390 $-$ 734 nm spectral window. Wavelength calibration is carried out using an iodine absorption cell, which also allows for characterisation of the instrumental profile. Spectra were reduced using the standard PFS reduction pipeline \citep{Butler:1996,Crane:2006} and radial velocity measurements were extracted using a custom IDL pipeline.

\subsubsection{CHIRON}

We obtained 15 spectra of TOI-4994 with the CHIRON high-resolution spectrograph on the SMARTS 1.5-m telescope located at CTIO in Chile. CHIRON is a fiber-fed echelle spectrograph with a resolving power of $R=80,000$ over the wavelength range $4100-8700$\,\AA{} \citep{tokovinin:2013}. The observations were taken between UT 2022 August 28 and October 24. Wavelength calibrations are provided by bracketing Thorium-Argon (Th-Ar) arc lamp exposures. Relative velocities were measured from each spectrum by deriving their stellar line broadening kernels via a Least Squares Deconvolution (LSD) analysis (described in \citealt{Zhou:2016}). 

\subsubsection{HARPS}

\begin{figure}[!ht]
\vspace{.1in}
\centering
\includegraphics[width=1\linewidth]{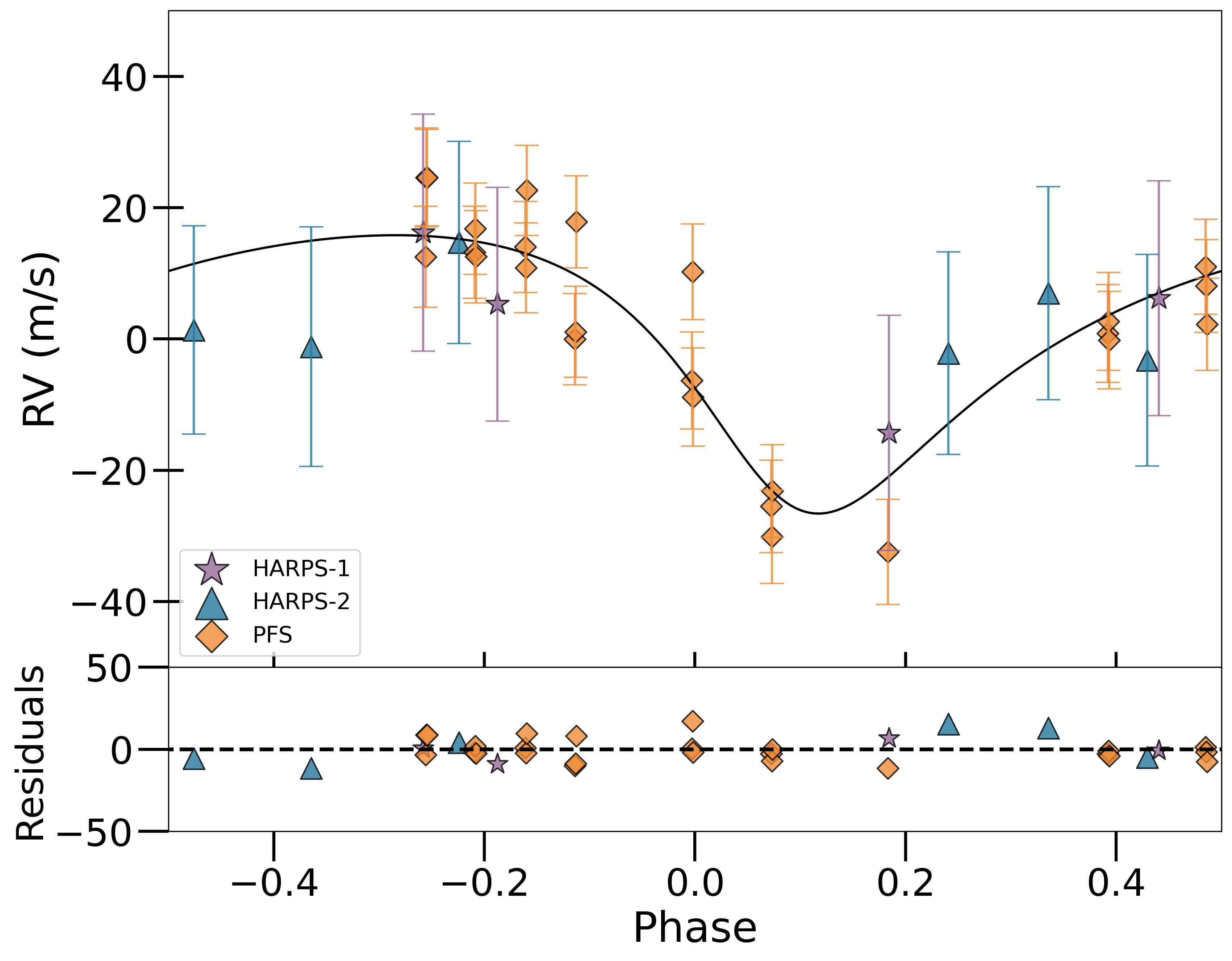}

\caption{Phase-folded radial velocities from PFS (golden diamonds) and HARPS (purple stars and blue triangles) only, with residuals plotted below.}
\label{fig:PFS_HARPS} 
\end{figure}

We obtained a total of 10 RV measurements from the High Accuracy Radial velocity Planet Searcher (HARPS; \citealp{Mayor:2003}) on the 3.6-m European Southern Observatory (ESO) telescope in La Silla Observatory, Chile. HARPS is a high-resolution spectrograph with a precision on the order of 1 m/s, a resolving power of 115,000 and a wavelength range of 378 $-$ 691 nm. The spectra and radial velocities were reduced with the HARPS dedicated pipeline. The HARPS observations of our target were taken by two independent teams and therefore reduced with different methodologies. We label these HARPS-1 and HARPS-2 to distinguish them. In Figure~\ref{fig:PFS_HARPS}, we plot only the most precise radial velocity data from PFS and HARPS to better highlight the best-fit RV model.


\subsubsection{CORALIE}

TOI-4994 was observed by the CORALIE fiber-fed high-resolution echelle spectrograph located at the Swiss 1.2-m Leonhard Euler Telescope at ESO’s La Silla Observatory \citep{Udry:2000, Queloz:2001}. It has a spectral resolution of 60,000 and a wavelength range of 390 $–$ 680 nm. A total of 7 RV observations of TOI-4994 were collected between UT 2022 April 11 and September 23. The observations have exposure times of 1800 seconds and a signal-to-noise ratio (S/N) ranging between 25 and 50 measured at 550 nm. The S/N variations are due to the different observing conditions. The RVs were obtained by cross-correlating each spectrum against a binary mask corresponding to the stellar type G2 (e.g., \citealt{Pepe:2002}). The resulting product is a cross-correlation function (CCF) from which the radial velocity is measured as well as several other indicators, such as the full width half maximum of the CCF, the bisector inverse slope and the contrast of the CCF. Additional indexes such as $H_\alpha$ and Ca indexes are also derived from each spectrum.

\subsubsection{FEROS}

Additionally, we obtained a total of 20 RV observations with the ESO's Fiber-fed Extended Range Optical Spectrograph \citep[FEROS][]{feros} located at La Silla Observatory in Chile. These observations were obtained in the context of the Warm gIaNts with tEss (WINE) collaboration, which focuses on the systematic discovery and characterization of transiting warm giant planets \citep[e.g.][]{brahm:2019,jordan:2020,brahm:2020,schlecker:2020,hobson:2021,trifonov:2021,trifonov:2023,brahm:2023,hobson:2023,eberhardt:2023}. FEROS has a spectral resolution of 48,000 and a wavelength coverage of 350 $-$ 920 nm. Observations of TOI-4994 were taken between UT 2022 July 3 and UT 2023 September 7, with exposure times of 1200 seconds and S/Ns between 30 and 50 per resolution element. FEROS data was processed with the \texttt{ceres} pipeline \citep{ceres}, which delivers precision radial velocities, bisector span measurements and a  rough estimation of the atmospheric stellar parameters.

\subsection{\rm \textbf{High-resolution Imaging with SOAR}}
\label{sec:soar}

\begin{figure}[!ht]
\vspace{.1in}
\centering
\includegraphics[width=1\linewidth]{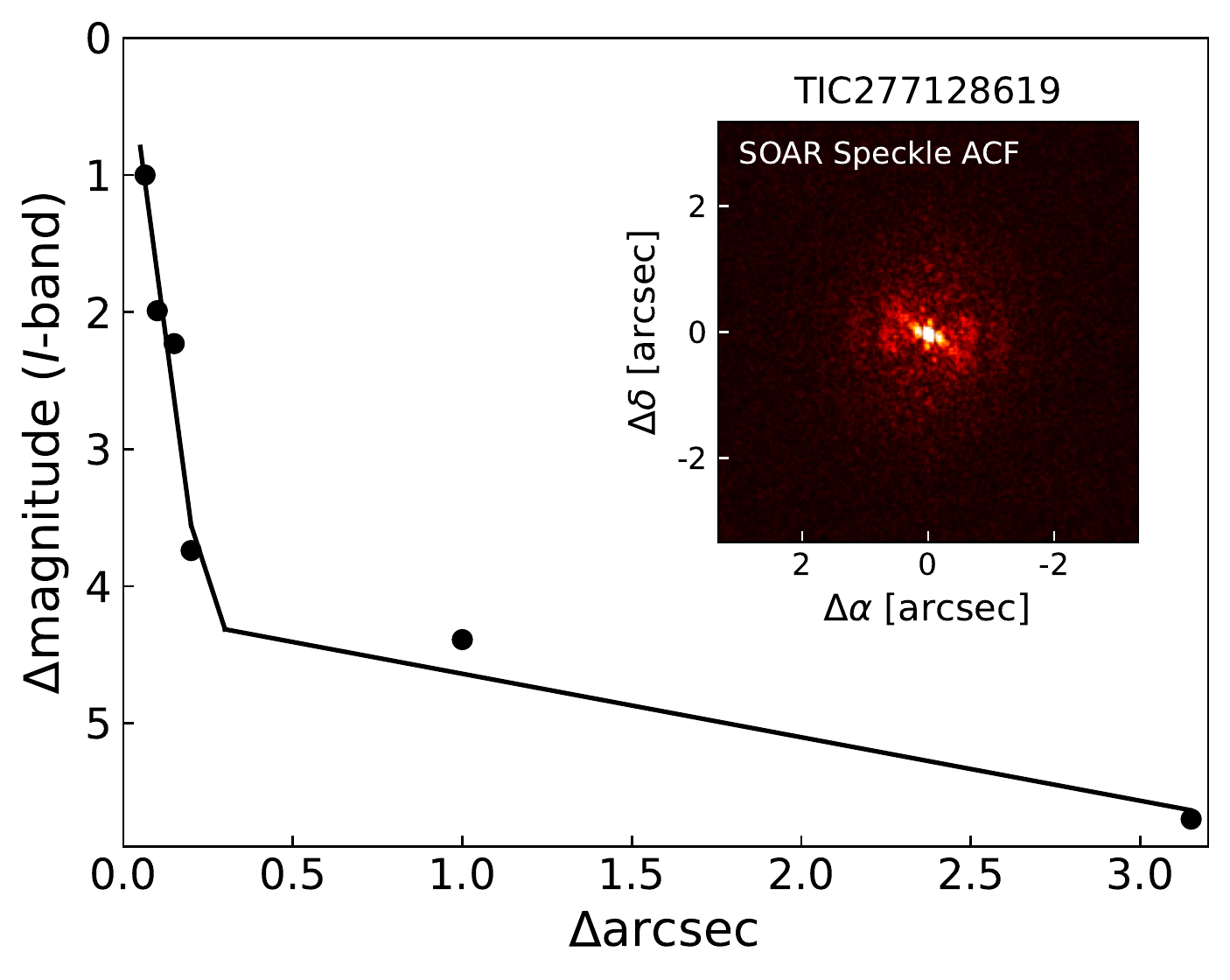}

\caption{$I$-band speckle image and sensitivity curve for TOI-4994 from the Southern Astrophysical Research (SOAR) telescope.}
\label{fig:SOAR} 
\end{figure}

We obtained high-resolution speckle images of TOI-4994 to check that the signals originate from the target star and not from another nearby stellar companion that could contaminate the SED and affect the inferred transit depth. The star was observed with the speckle camera at the Southern Astrophysical Research (SOAR) 4.1-m telescope on UT 2022 August 18. The data reduction process, as well as a thorough description of the SOAR-TESS survey can be found in \citet{Tokovinin:2018} and in \citet{ziegler:2019}. In Figure~\ref{fig:SOAR}, we show the 5$\sigma$ limit of companion detection, and the inset shows the speckle auto-correlation function (ACF). From this observation, we can exclude the presence of any stellar companion within about 4.5 magnitudes of the brightness of TOI-4994 and an angular separation between 0.25$\arcsec$ and 2$\arcsec$, and within about 5 magnitudes for larger separations.

\begin{figure}
	\centering\vspace{.0in}
	\includegraphics[width=1\linewidth]{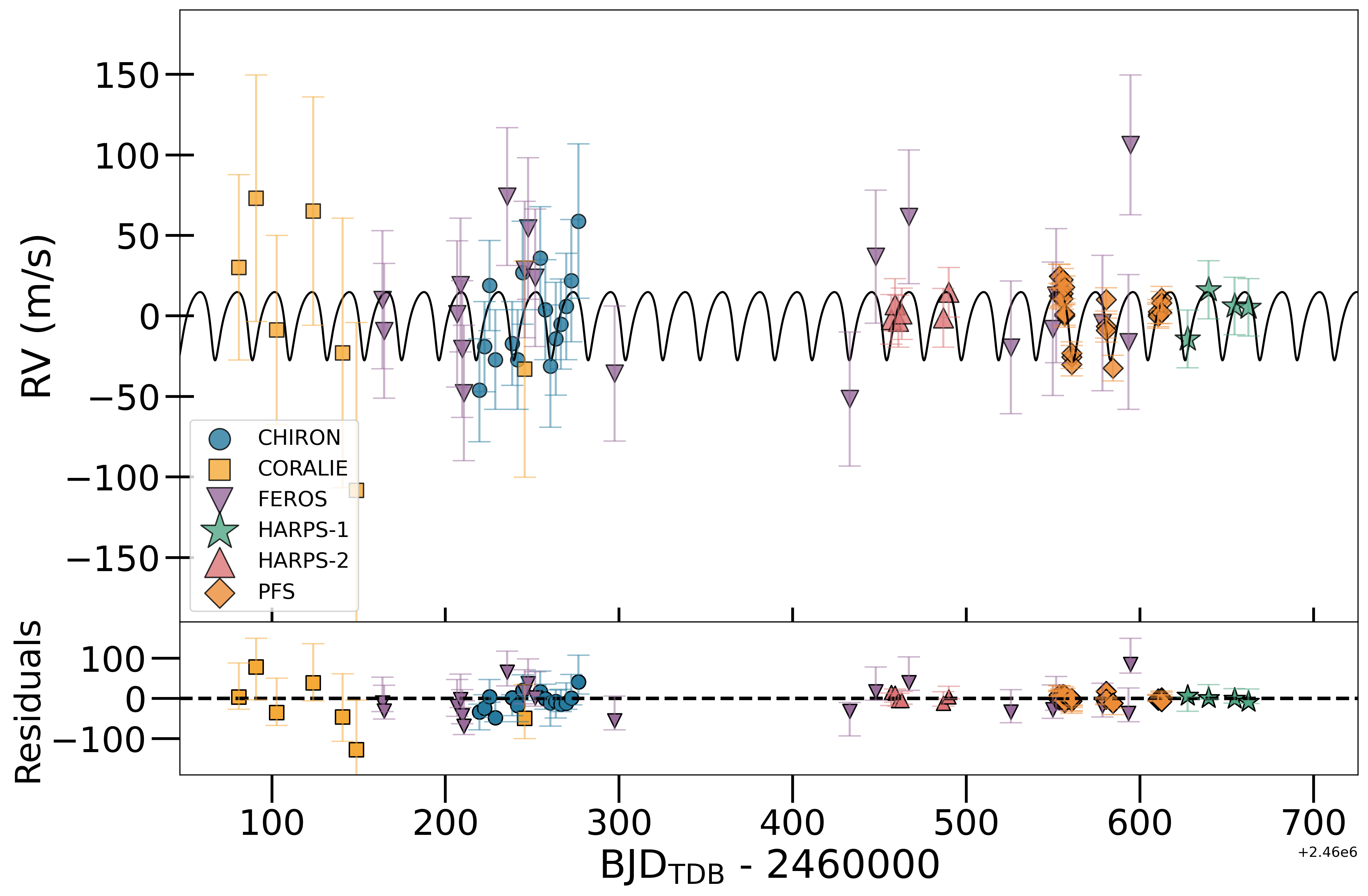}
	\includegraphics[width=1\linewidth]{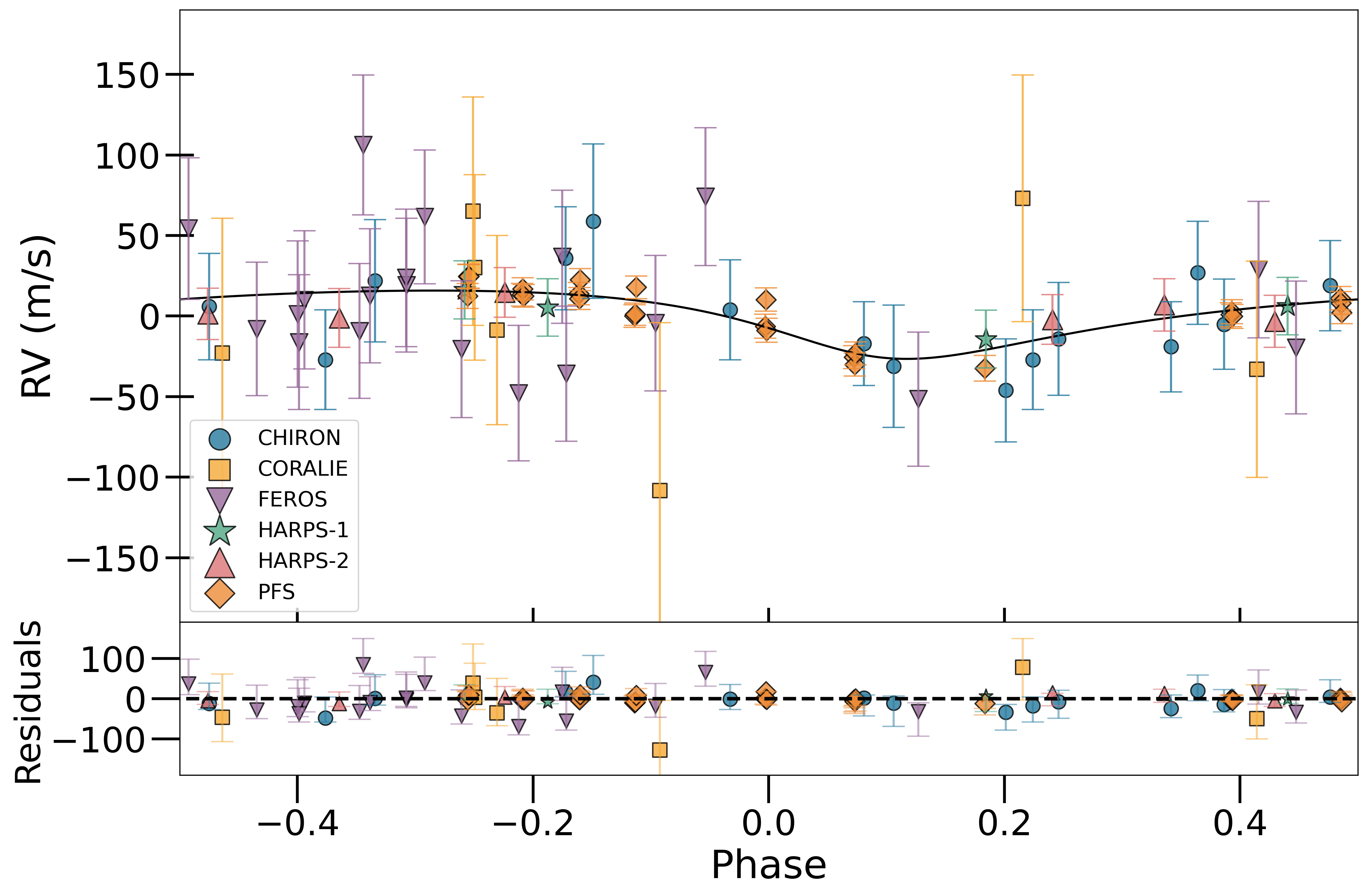}
	\caption{Radial velocity measurements of TOI-4994 \textbf{(top)} phase-folded on the orbital period \textbf{(bottom)}. The black lines show the best-fit {\tt EXOFASTv2} models, with the residuals shown below the RV curves.}
	\label{fig:RVs} 
\end{figure}

\section{Global Analysis} \label{sec:analysis}

To constrain the stellar and planetary parameters of the system, we conducted a global fit of all our data using the publicly available exoplanet modeling suite, {\tt EXOFASTv2} \citep{Eastman:2013,Eastman:2017,eastman:2019}. The physical properties of the host star were determined from a combination of the spectral energy distribution (SED) fit, as well as spectroscopic priors from our CHIRON data (including an initial estimate of the effective temperature and $\log{g}$), a parallax prior from \textit{Gaia} DR3 \citep{Gaia:2023}, and an upper bound on the extinction from galactic dust maps. For the SED fit, we incorporated available photometry from \textit{Gaia} $G$, $G_{BP}$, and $G_{RP}$ bands, 2MASS $J$, $H$, and $K$ magnitudes, and WISE $W_{1}-W_{4}$ magnitudes 
(see Table~\ref{tbl:LitProps} for these values). 
A \citet{Kurucz:1992} atmosphere model was fit to all the available fluxes. As priors to the SED and global fit, we adopted a parallax of $\mu = 3.022^{+0.022}_{-0.023}$ mas from \textit{Gaia} EDR3 (corrected as per \citealt{Lindegren:2021}) 
We set an upper limit on the maximum line-of-sight visual extinction $A_{V}$ of 0.350 from the \citet{Schlegel:1998} extinction maps. We used a prior on [Fe/H] of $0.11 \pm 0.09$ dex, which is the median metallicity of all our spectra from CHIRON. 

\begin{figure}[!ht]
\vspace{.1in}
\centering
\includegraphics[width=1\linewidth]{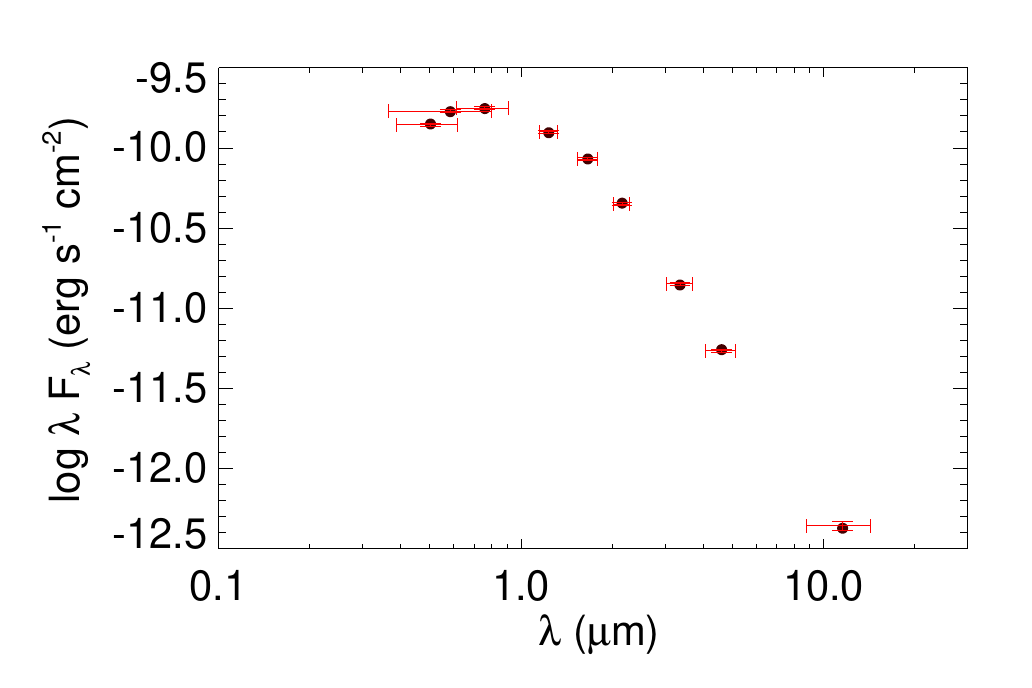}

\caption{Spectral Energy Distribution (SED) fit for TOI-4994. The horizontal error bars correspond to the bandwidth of each filter, while the error bars in flux are the measurement uncertainties. The black points are the modeled, broadband averages.}
\label{fig:sed} 
\end{figure}

\begin{figure}[!ht]
\vspace{.1in}
\centering
\includegraphics[width=1\linewidth]{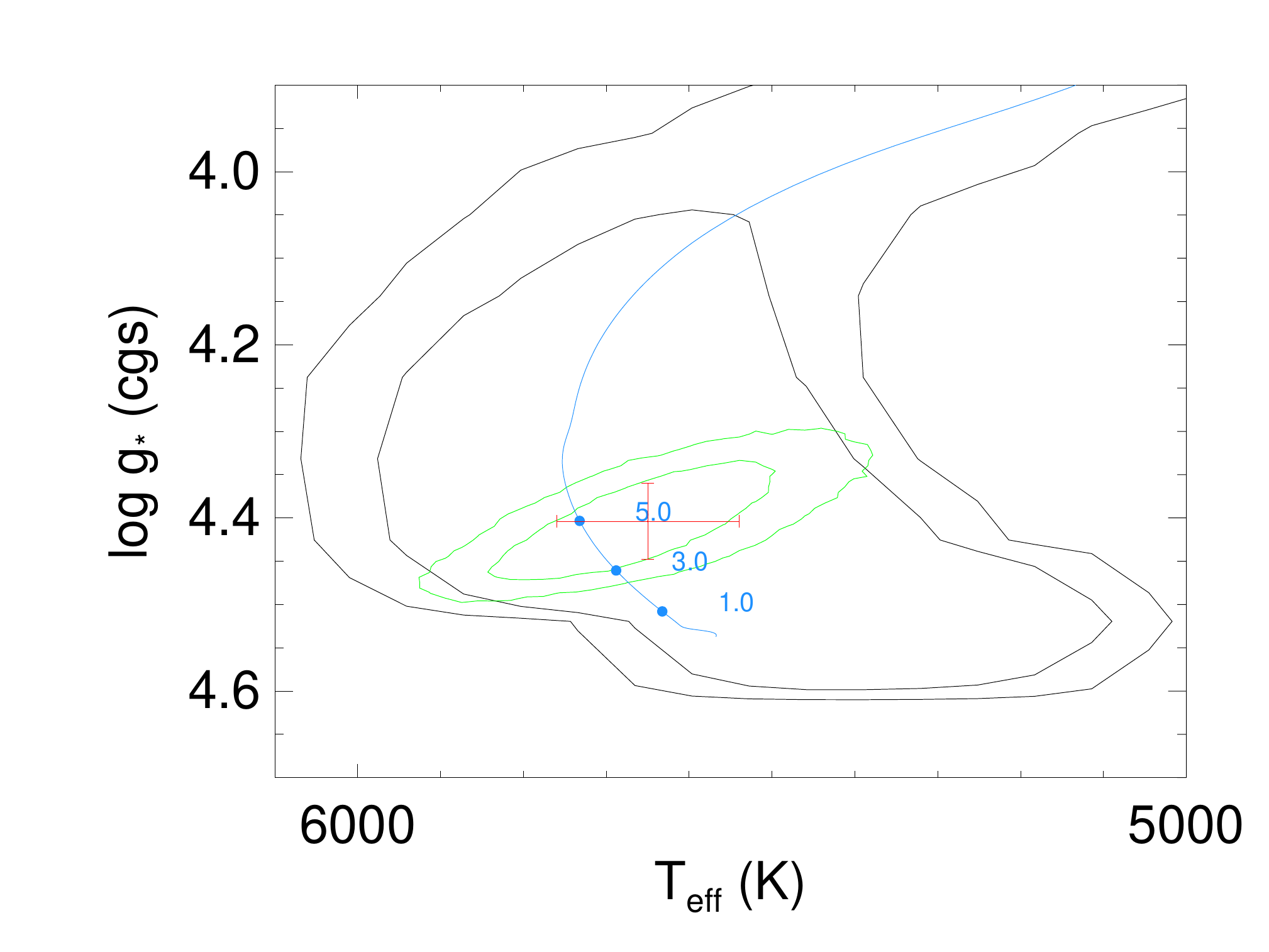}

\caption{Evolutionary track and present evolutionary stage of TOI-4994 based on the best-fit MIST model. The blue line is the MIST track and the black contours represent the 1$\sigma$ and 2$\sigma$ constraints on the current temperature and surface gravity from the MIST isochrone, while the green contours show the 1$\sigma$ and 2$\sigma$ constraints on $T_{\rm eff}$ and log$g_{*}$ from the {\tt EXOFASTv2} fit and observations of the system. The red cross indicates the median value and confidence intervals.}
\label{fig:MIST} 
\end{figure}

Within this SED fit, we use the MESA Isochrones and Stellar Tracks (MIST) stellar evolution models \citep{Dotter:2016}. Figure~\ref{fig:sed} illustrates the resulting SED fit, while the MIST evolutionary track is shown in Figure \ref{fig:MIST}. The inferred stellar parameters are summarized in Table~\ref{tab:stellar}.

We infer a stellar mass of $M_{\star} = 1.005^{+0.064}_{-0.061}M_{\odot}$, a radius of $R_{\star}= 1.055^{+0.040}_{-0.037} R_{\odot}$, and an effective temperature of $T_{\rm eff} =5640 \pm 100$ K, making this host a middle-aged ($6.3^{+4.2}_{-3.8}$ Gyr) G5 dwarf (per the classifications by \citealt{Pecaut:2013}) and an almost identical solar twin. The star exhibits a slightly enhanced metallicity, however, with [Fe/H] = $0.165^{+0.083}_{-0.084}$ dex. Using the empirical $R'_{\rm HK}$-age relations by \citet{Mamajek:2008}, we derived an age for the system of 7.8 $\pm$ 1.3 Gyr, which is consistent with the age from our SED fit. These relations also predict a stellar rotation period of  46.2 $\pm$ 3.1 days.

The planetary parameters were determined through a simultaneous global fit of the system, which combined the TESS and ground-based photometry and the six RV datasets described in Sections~\ref{sec:TESS_photometry},~\ref{ground_based}, and ~\ref{sec:spectroscopy}. {\tt EXOFASTv2} determines the stellar and planetary parameters using a differential evolution Markov Chain Monte Carlo method (see \citealt{Eastman:2013} for further details). We adopted an initial guess on the ephemeris from an independent fit to the TESS light curves. The resulting planetary parameters are listed in Table~\ref{tab:planetary}.

\section{Results and Discussion} 
\label{sec:results_discussion}

From our global fit to the available photometry and radial velocity datasets, we determined a planetary mass of $M_{P}= 0.280^{+0.037}_{-0.034} M_{J}$ (or $89.1~M_{\oplus}$), a radius of $R_{P}= 0.762^{+0.030}_{-0.027}R_{J}$ (or $8.55 R_{\oplus}$), and a bulk density of $\rho_{p} = 0.78^{+0.16}_{-0.14}$ $\rm g/cm^3$. These properties are similar to those of Saturn, and, given the planet's orbital period, it fits in with the population of warm Saturns. The number of well-characterized warm Saturns in the literature is still small. There are currently less than 20 planets between 0.8--1.1$M_{Sat}$ and with periods between 10--200 days that have directly measured masses and radii,  making this object a valuable addition to this rare sample. Some of these include HD 89345b \citep{VanEylen:2018}, K2-287b \citep{Jordan:2019}, K2-261b \citep{Johnson:2018}, NGTS-11b (TOI-1847b; \citealt{Gill:2020}), and K2-232b \citep{brahm:2020}.

\begin{figure*}[!ht]
\begin{center}
    \includegraphics[width=0.75\textwidth]{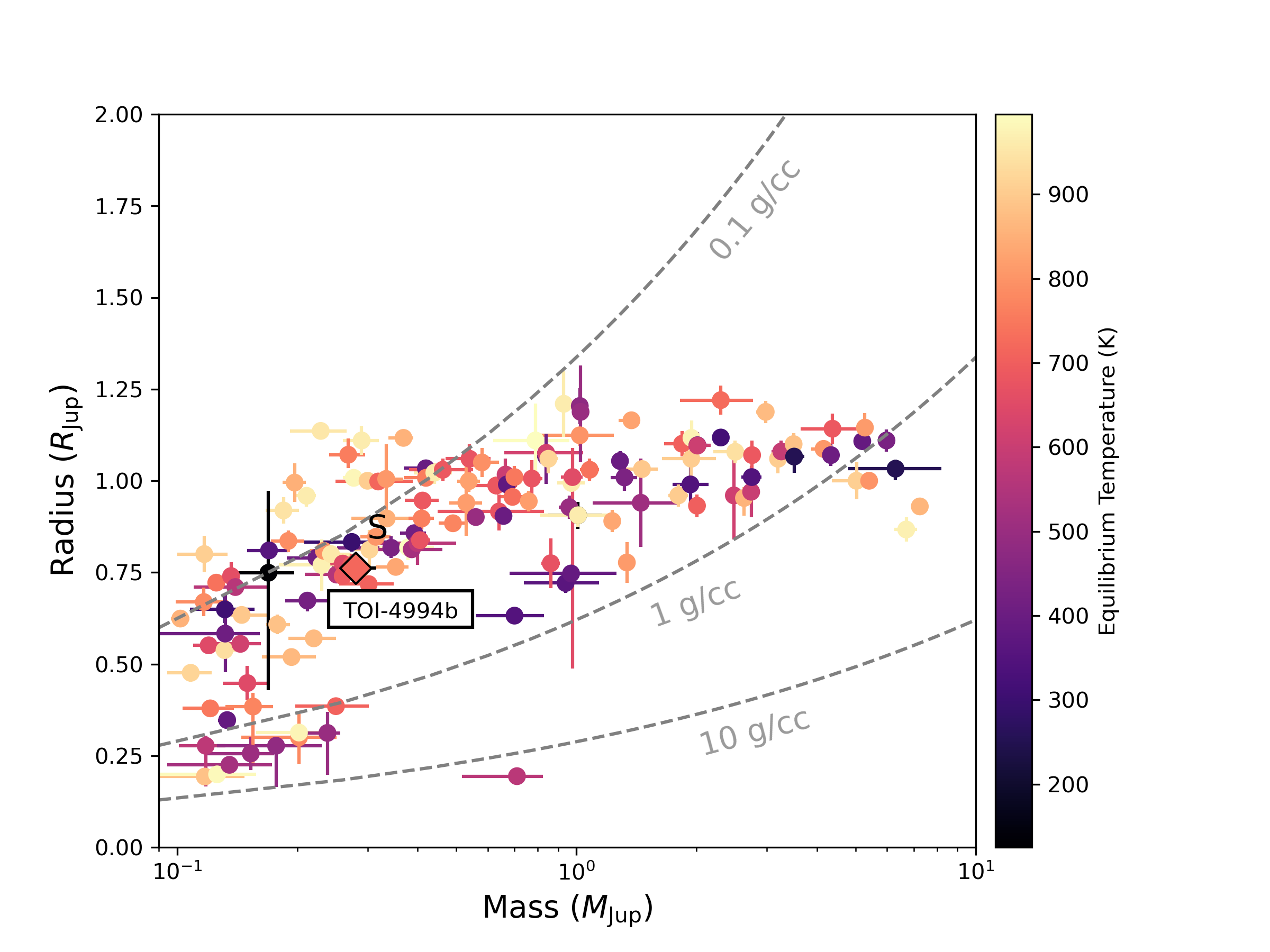}
\end{center}
\caption{Mass-radius diagram of all warm giants (planet masses between 0.1 $\leq M_{\rm Jup} \leq $ 10 and temperatures below 1000 K). The grey dashed lines are lines of constant bulk density. TOI-4994b is represented by a diamond, and Saturn is over-plotted for reference.}

\label{fig:MRdiagram_all}
\end{figure*}

Although the mass and radius of TOI-4994b are slightly too large to categorize it as a sub-Saturn, (typically defined as planets with $4 < R_p/R_{\oplus} < 8$) it shares many characteristics with this population, described in detail by \citet{Petigura:2017}. Notably, TOI-4994b is a single planet (with no evidence of other companions in the system), and has an eccentric orbit ($e= 0.32^{+0.048}_{-0.047}$). Such high eccentricity and lack of observed companions (characteristic of the most massive sub-Saturns) may be evidence of past dynamical instability, as hypothesized by \citet{Petigura:2017}.  In addition, TOI-4994b orbits a metal-rich star ([Fe/H] = $0.165^{+0.083}_{-0.084}$ dex), like the majority of known sub-Saturns. Interestingly, \citet{Petigura:2017} also found strong correlations between planet mass, eccentricity, and host star metallicity.

Fairnington et al. (in prep) identified a potential bimodality in the eccentricity distribution of warm, single sub-Saturns, noting a scarcity of moderately eccentric planets ($e = 0.2 - 0.5$). TOI-4994b is the third confirmed object in this range, and its properties suggest a history shaped by planet-planet scattering and a possible merger (\citealt{Petigura:2017,Brady:2018,Millholland:2022}). If a merger were the origin history of this planet, it may have a high bulk metallicity, which could be probed with JWST. While TOI-4994b’s eccentricity and orbital distance are inconsistent with high eccentricity migration, additional radial velocity follow-up to search for long-period non-transiting companions may provide additional insights into its formation history.

\begin{table*}
\scriptsize
\centering
\caption{Median values and 68\% confidence interval for global model of TOI-4994}
\label{tab:stellar}

\begin{tabular}{llcccc}
  \hline
  \hline
Parameter & Units & Values & & & \\
\hline
~~~~$M_*$\dotfill &Mass (\msun)\dotfill &$1.005^{+0.064}_{-0.061}$\\
~~~~$R_*$\dotfill &Radius (\rsun)\dotfill &$1.055^{+0.040}_{-0.037}$\\
~~~~$L_*$\dotfill &Luminosity (\lsun)\dotfill &$1.020^{+0.041}_{-0.052}$\\
~~~~$F_{Bol}$\dotfill &Bolometric Flux (cgs)\dotfill &($2.98^{+0.11}_{-0.15}) \times 10^{-10}$\\
~~~~$\rho_*$\dotfill &Density (cgs)\dotfill &$1.21^{+0.17}_{-0.16}$\\
~~~~$\log{g}$\dotfill &Surface gravity (cgs)\dotfill &$4.393^{+0.045}_{-0.046}$\\
~~~~$T_{\rm eff}$\dotfill &Effective Temperature (K)\dotfill &$5640\pm110$\\
~~~~$[{\rm Fe/H}]$\dotfill &Metallicity (dex)\dotfill &$0.165^{+0.083}_{-0.084}$\\
~~~~$[{\rm Fe/H}]_{0}$\dotfill &Initial Metallicity$^{1}$ \dotfill &$0.176^{+0.079}_{-0.080}$\\
~~~~$Age$\dotfill &Age (Gyr)\dotfill &$6.3^{+4.2}_{-3.8}$\\
~~~~$EEP$\dotfill &Equal Evolutionary Phase$^{2}$ \dotfill &$380^{+31}_{-41}$\\
~~~~$A_V$\dotfill &V-band extinction (mag)\dotfill &$0.295^{+0.041}_{-0.073}$\\
~~~~$\sigma_{SED}$\dotfill &SED photometry error scaling \dotfill &$0.73^{+0.31}_{-0.19}$\\
~~~~$\varpi$\dotfill &Parallax (mas)\dotfill &$3.022^{+0.022}_{-0.023}$\\
~~~~$d$\dotfill &Distance (pc)\dotfill &$330.9^{+2.5}_{-2.4}$\\
  \hline

\end{tabular}
\begin{flushleft} 
  \footnotesize{ 
    \textbf{\textsc{NOTES:}}
\tablenotetext{}{See Table 3 in \citet{eastman:2019} for a detailed description of all parameters.}
\tablenotetext{1}{The metallicity of the star at birth.}
\tablenotetext{2}{Corresponds to static points in a star's evolutionary history. See \S2 in \citet{Dotter:2016}.}
             }
             
 \end{flushleft}
\end{table*}
\begin{table*}
\scriptsize
\centering
\caption{Median values and 68\% confidence interval for global model of TOI-4994 b}
\label{tab:planetary}
\begin{tabular}{llcccc}
  \hline
  \hline
Parameter & Description (Units) & Values & & & \\
~~~~$P$\dotfill &Period (days)\dotfill &$21.491984\pm0.000023$\\
~~~~$R_P$\dotfill &Radius (\rj)\dotfill &$0.762^{+0.030}_{-0.027}$\\
~~~~$M_P$\dotfill &Mass (\mj)\dotfill &$0.280^{+0.037}_{-0.034}$\\
~~~~$T_C$\dotfill &Time of conjunction$^{4}$ (\bjdtdb)\dotfill &$2459363.69480\pm0.00076$\\
~~~~$T_T$\dotfill &Time of minimum projected separation$^{5}$ (\bjdtdb)\dotfill &$2459363.69484\pm0.00076$\\
~~~~$T_0$\dotfill &Optimal conjunction Time$^{6}$ (\bjdtdb)\dotfill &$2459965.47035^{+0.00042}_{-0.00041}$\\
~~~~$a$\dotfill &Semi-major axis (AU)\dotfill &$0.1515^{+0.0032}_{-0.0031}$\\
~~~~$i$\dotfill &Inclination (Degrees)\dotfill &$89.56^{+0.29}_{-0.32}$\\
~~~~$e$\dotfill &Eccentricity \dotfill &$0.319^{+0.048}_{-0.047}$\\
~~~~$\omega_*$\dotfill &Argument of Periastron (Degrees)\dotfill &$143.1^{+9.5}_{-10.}$\\
~~~~$T_{eq}$\dotfill &Equilibrium temperature$^{7}$ (K)\dotfill &$717.6^{+9.7}_{-10.}$\\
~~~~$\tau_{\rm circ}$$^\dagger$\dotfill &Tidal circularization timescale (Gyr)\dotfill &$610^{+400}_{-260}$\\
~~~~$K$\dotfill &RV semi-amplitude (m/s)\dotfill &$21.6^{+2.3}_{-2.4}$\\
~~~~$R_P/R_*$\dotfill &Radius of planet in stellar radii \dotfill &$0.07424^{+0.00078}_{-0.00068}$\\
~~~~$a/R_*$\dotfill &Semi-major axis in stellar radii \dotfill &$30.9\pm1.4$\\
~~~~$\delta$\dotfill &$\left(R_P/R_*\right)^2$ \dotfill &$0.00551^{+0.00012}_{-0.00010}$\\
~~~~$\delta_{\rm B}$\dotfill &Transit depth in B (fraction)\dotfill &$0.00829^{+0.00039}_{-0.00035}$\\
~~~~$\delta_{\rm R}$\dotfill &Transit depth in R (fraction)\dotfill &$0.00684^{+0.00025}_{-0.00024}$\\
~~~~$\delta_{\rm g'}$\dotfill &Transit depth in g' (fraction)\dotfill &$0.00788^{+0.00026}_{-0.00024}$\\
~~~~$\delta_{\rm i'}$\dotfill &Transit depth in i' (fraction)\dotfill &$0.00662\pm0.00016$\\
~~~~$\delta_{\rm TESS}$\dotfill &Transit depth in TESS (fraction)\dotfill &$0.00657\pm0.00013$\\
~~~~$\delta_{\rm V}$\dotfill &Transit depth in V (fraction)\dotfill &$0.00708^{+0.00025}_{-0.00024}$\\
~~~~$\tau$\dotfill &Ingress/egress transit duration (days)\dotfill &$0.01319^{+0.0011}_{-0.00041}$\\
~~~~$T_{14}$\dotfill &Total transit duration (days)\dotfill &$0.1855^{+0.0013}_{-0.0011}$\\
~~~~$T_{FWHM}$\dotfill &FWHM transit duration (days)\dotfill &$0.17213^{+0.00092}_{-0.00091}$\\
~~~~$b$\dotfill &Transit Impact parameter \dotfill &$0.18^{+0.14}_{-0.12}$\\
~~~~$b_S$\dotfill &Eclipse impact parameter \dotfill &$0.26^{+0.18}_{-0.17}$\\
~~~~$\tau_S$\dotfill &Ingress/egress eclipse duration (days)\dotfill &$0.0202^{+0.0024}_{-0.0019}$\\
~~~~$T_{S,14}$\dotfill &Total eclipse duration (days)\dotfill &$0.265^{+0.029}_{-0.025}$\\
~~~~$T_{S,FWHM}$\dotfill &FWHM eclipse duration (days)\dotfill &$0.245^{+0.028}_{-0.025}$\\
~~~~$\delta_{S,2.5\mu m}$$^\dagger$\dotfill &Blackbody eclipse depth at 2.5$\mu$m (ppm)\dotfill &$3.22^{+0.39}_{-0.35}$\\
~~~~$\delta_{S,5.0\mu m}$$^\dagger$\dotfill &Blackbody eclipse depth at 5.0$\mu$m (ppm)\dotfill &$67.8^{+4.5}_{-4.1}$\\
~~~~$\delta_{S,7.5\mu m}$$^\dagger$\dotfill &Blackbody eclipse depth at 7.5$\mu$m (ppm)\dotfill &$165.5^{+8.4}_{-7.7}$\\
~~~~$\rho_P$\dotfill &Density (cgs)\dotfill &$0.78^{+0.16}_{-0.14}$\\
~~~~$logg_P$\dotfill &Surface gravity \dotfill &$3.077^{+0.069}_{-0.074}$\\
~~~~$\fave$\dotfill &Incident Flux (\fluxcgs)\dotfill &$0.0544^{+0.0029}_{-0.0031}$\\
~~~~$T_P$\dotfill &Time of Periastron (\bjdtdb)\dotfill &$2459343.85^{+0.38}_{-0.35}$\\
~~~~$T_S$\dotfill &Time of eclipse (\bjdtdb)\dotfill &$2459370.99^{+0.72}_{-0.74}$\\
~~~~$V_c/V_e$\dotfill & \dotfill &$0.797\pm0.039$\\
~~~~$e\cos{\omega_*}$\dotfill & \dotfill &$-0.250^{+0.053}_{-0.056}$\\
~~~~$e\sin{\omega_*}$\dotfill & \dotfill &$0.189^{+0.048}_{-0.047}$\\
\hline
\end{tabular}
 \begin{flushleft} 
  \footnotesize{ 
    \textbf{\textsc{\hspace{0.75in}NOTES:}}
\tablenotetext{}{See Table 3 in \citet{eastman:2019} for a detailed description of all parameters.}
$^\dagger$ These values are predicted values from theory and not directly observed. 
\tablenotetext{4}{Time of conjunction is commonly reported as the ``transit time".}
\tablenotetext{5}{Time of minimum projected separation is a more correct ``transit time".}
\tablenotetext{6}{Optimal time of conjunction minimizes the covariance between $T_C$ and Period.}
\tablenotetext{7}{Assumes no albedo and perfect redistribution.}
               }
 \end{flushleft}
\end{table*}

TOI-4994b has a relatively high equilibrium temperature of $717.6^{+9.7}_{-10}$ K, calculated using Equation 1 from \citet{Hansen:2007} within {\tt EXOFASTv2}:

\begin{equation}
    T_{eq} = T_{\rm eff}\sqrt{\frac{R_{\star}}{2a}},
\end{equation}

and which assumes an albedo of zero and perfect heat redistribution. We also estimate an orbit-averaged incident flux of $5.4\times10^7$ $\rm erg~s^{-1} cm^{-2}$, using the following formula: 

\begin{equation}
    \bigl \langle F \bigr \rangle= \sigma_{B}T_{\rm eff}^{4}\bigg(\frac{R_{\star}}{a}\bigg(1-e^2/2\bigg )\bigg)^{2},
\end{equation}

and we find that TOI-4994b is below the \citet{Demory:2011} insolation threshold of $2\times10^{8}$ $\rm erg$ $\rm s^{-1} \rm cm^{-2}$. Figure~\ref{fig:MRdiagram_all} highlights the mass and radius of TOI-4994b within the population of warm giant planets in the NASA Exoplanet Archive \citep{Akeson:2013}. The planets are color-coded by their reported equilibrium temperature.

The atmospheric composition of a planet provides key insights into its formation, history, and potential for habitability (see, e.g., \citealt{Madhusudhan:2019}). In order to assess the prospects for follow-up atmospheric characterization of TOI-4994b with JWST, we evaluated its Transmission Spectroscopy Metric (TSM; \citealt{Kempton:2018}). This value is proportional to the expected signal-to-noise ratio of a transmission spectrum of the target, with higher numbers indicating better prospects for observation. We obtained a TSM value of 30 for TOI-4994b, which is below the value of 90 recommended by \citet{Kempton:2018} for planets between 4$-$10$R_{\oplus}$. 

Finally, we consider the feasibility of measuring the spin-orbit misalignment of the system via the Rossiter-McLaughlin (RM) effect or Doppler Tomography. We estimate an expected RM signal of 8.3 m/s for TOI-4994b using Equation 12 of \citet{Albrecht:2022}, assuming a rotational velocity of \vsini = 2.2 km/s, which is a typical value for G-type stars (e.g., \citealt{dosSantos:2016}). Alternatively, using the predicted stellar rotation rate of 46.2 $\pm$ 3.1 days estimated in Section~\ref{sec:analysis}, we derived an RM signal of 4.3 m/s.  Although TOI-4994 is slightly fainter than other stars hosting a warm giant, this signal is potentially observable with high-resolution instruments, such as PFS, and we therefore encourage future follow-up observations to determine the system's spin-orbit alignment.

\section{Conclusions} 
\label{sec:conclusions}

We presented the discovery and characterization of TOI-4994b, a warm Saturn transiting a moderately bright (V = 12.6 mag) G-type star in an eccentric, 21.5-day orbit, discovered by TESS. We confirmed the planetary nature of TOI-4994 with extensive ground-based photometry and with multiple high-resolution radial velocities from CHIRON, PFS, HARPS, FEROS, and CORALIE, and we ruled out the presence of nearby stellar companions with SOAR images. From our observations, we determined that TOI-4994 has physical properties similar to Saturn's, and consistent with the population of known sub-Saturns. We encourage further observations of this system, as it is a promising is candidate for spin-orbit angle measurements via the Rossiter-MacLaughlin effect. Such observations could shed more light on the formation and evolution of these rare planets.

\begin{center}
    ACKNOWLEDGEMENTS
\end{center}
RRM thanks Samuel Yee for conversations that improved this manuscript. T.T. acknowledges support by the BNSF program "VIHREN-2021" project No. KP-06-DV/5. This paper includes data collected by the TESS mission. Funding for the TESS mission is provided by the NASA's Science Mission Directorate. This research has also made use of the Exoplanet Follow-up Observation Program website, which is operated by the California Institute of Technology, under contract with the National Aeronautics and Space Administration under the Exoplanet Exploration Program. This paper includes data gathered with the 6.5-meter Magellan Telescopes located at Las Campanas Observatory, Chile. Additionally, some observations were collected at the European Southern Observatory under ESO programmes 111.250B.001 and 112.25W1.001. ML acknowledges support of the Swiss National Science Foundation under grant number PCEFP2\_194576.

This work makes use of observations from the LCOGT network. Part of the LCOGT telescope time was granted by NOIRLab through the Mid-Scale Innovations Program (MSIP). MSIP is funded by NSF.
Funding for the TESS mission is provided by NASA's Science Mission Directorate. 
This work has made use of data from the European Space Agency (ESA) mission
{\it Gaia} (\url{https://www.cosmos.esa.int/gaia}), processed by the {\it Gaia}
Data Processing and Analysis Consortium (DPAC,\url{https://www.cosmos.esa.int/web/gaia/dpac/consortium}). The contributions of ML, SUM, MB, AP, PF, FB, and SU were carried out within the framework of the NCCR PlanetS supported by the Swiss National Science Foundation under grants $51NF40\_182901$ and $51NF40\_205606$.The postdoctoral fellowship of KB is funded by F.R.S.-FNRS grant T.0109.20 and by the Francqui Foundation. The results reported herein benefitted from collaborations and/or information exchange within NASA’s Nexus for Exoplanet System Science (NExSS) research coordination network sponsored by NASA’s Science Mission Directorate under Agreement No. 80NSSC21K0593 for the program ``Alien Earths".  We acknowledge the use of public TESS data from pipelines at the TESS Science Office and at the TESS Science Processing Operations Center. Resources supporting this work were provided by the NASA High-End Computing (HEC) Program through the NASA Advanced Supercomputing (NAS) Division at Ames Research Center for the production of the SPOC data products.


Funding for the DPAC
has been provided by national institutions, in particular the institutions
participating in the {\it Gaia} Multilateral Agreement. Finally, this research has made use of the NASA Exoplanet Archive, which is operated by the California Institute of Technology, under contract with the National Aeronautics and Space Administration under the Exoplanet Exploration Program. KAC acknowledges support from the TESS mission via subaward s3449 from MIT. \\

\vspace{5mm}
\facilities{TESS, LCOGT, ASTEP-Antarctica, PFS, HARPS, CHIRON, CORALIE, FEROS, SOAR, Exoplanet Archive.}

\software{{AstroImageJ \citep{Collins:2017}, \tt EXOFASTv2} \citep{eastman:2019}, numpy \citep{harris:2020}, TAPIR \citep{Jensen:2013}.}

\appendix

\begin{figure*}[!ht]
    \centering
    \begin{tabular}{ccc} 
        \includegraphics[width=0.5\textwidth]{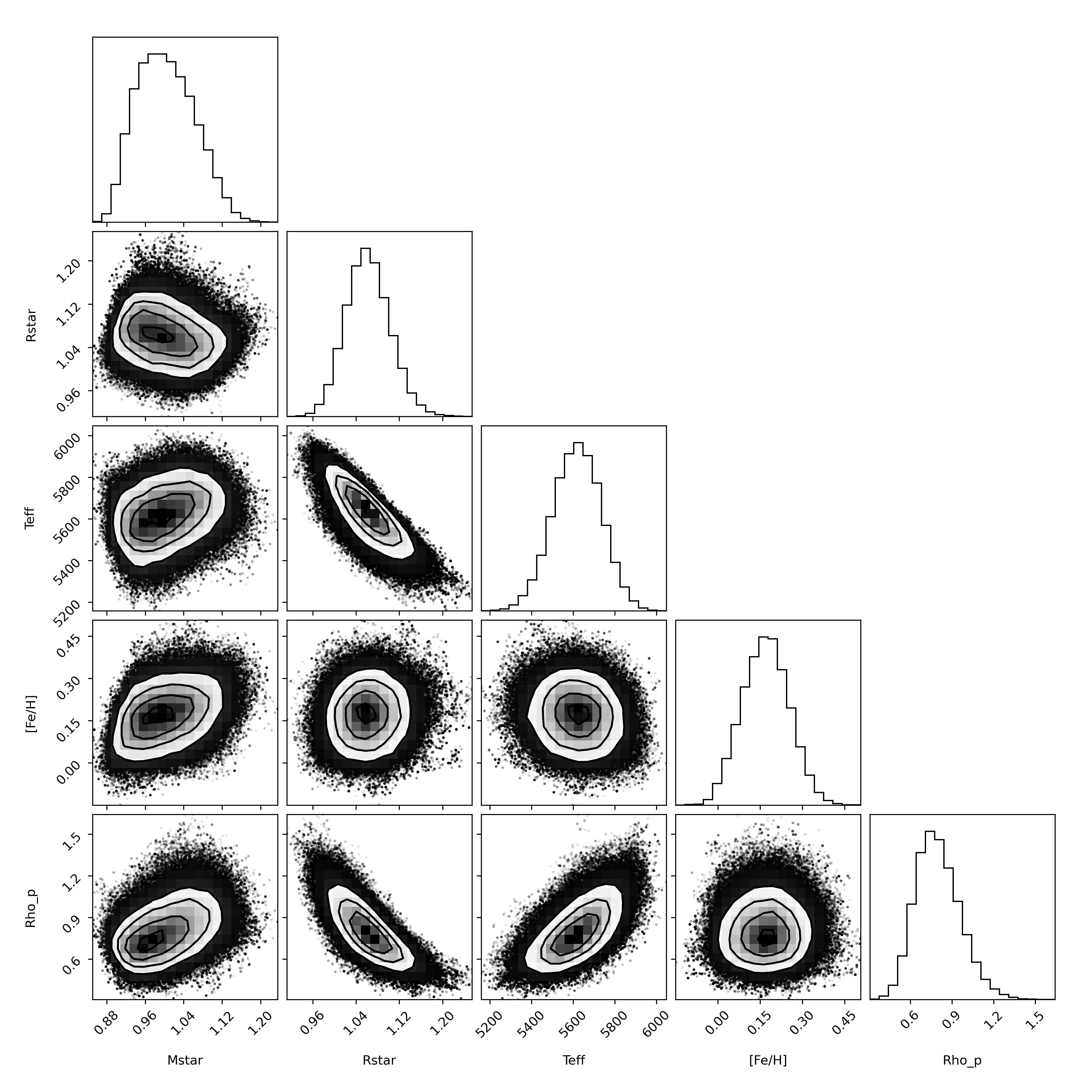} &
        \includegraphics[width=0.5\textwidth]{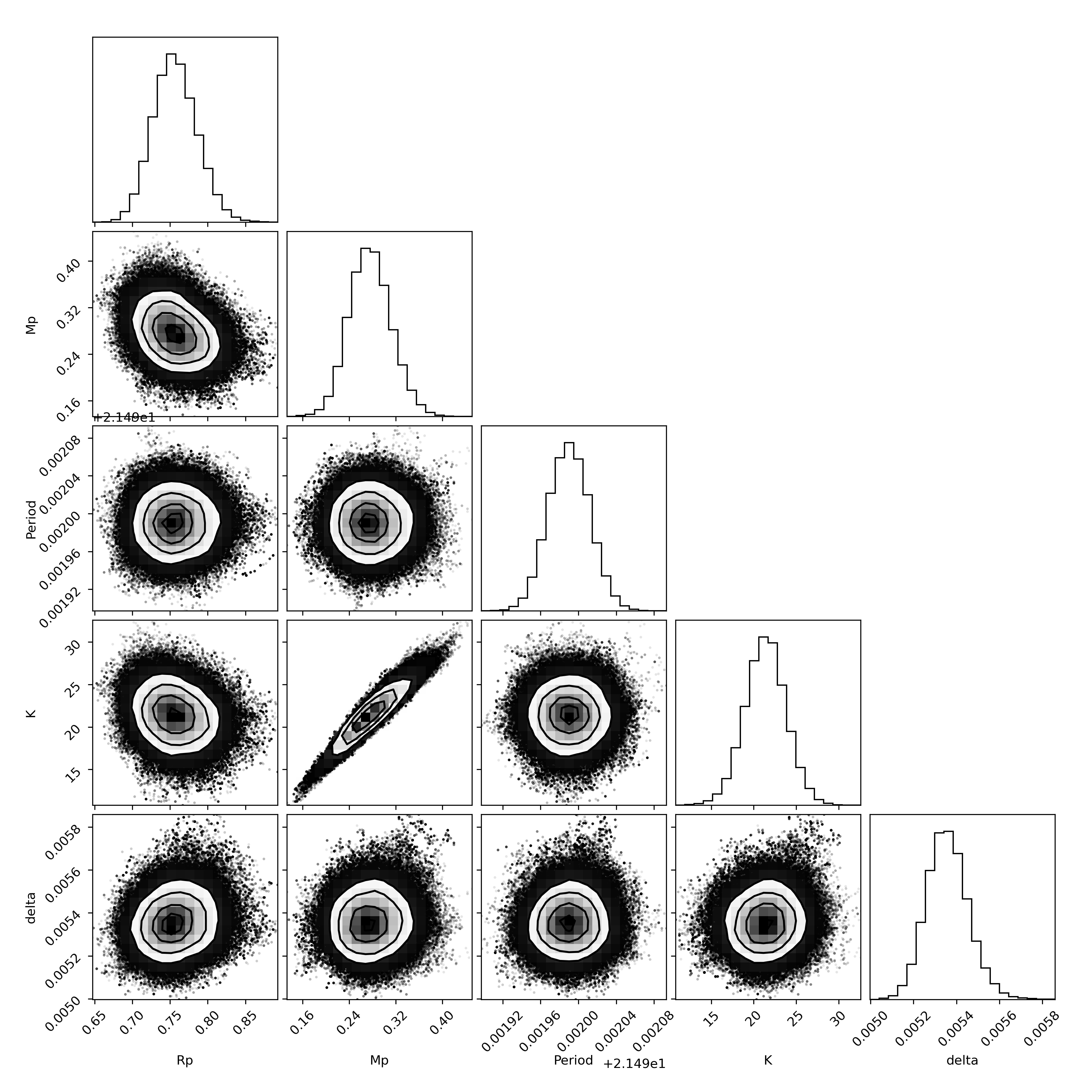}\\
    \end{tabular}
    \caption{Corner plots showing the covariances for some of the stellar and planetary parameters in our analysis.}
    \label{fig:corner_plots}
\end{figure*}

\begin{deluxetable*}{lccc}
\caption{Radial Velocity Data}
\label{tab:RVs}
\tablehead{
\colhead{Time (BJD)} & \colhead{RV (m/s)} & \colhead{RV error (m/s)} & \colhead{Instrument}}
\startdata
2460153.61049 & 4.40   & 4.13 & Planet Finder Spectrograph\\
2460153.62499 & 16.50  & 3.84 & Planet Finder Spectrograph\\
2460153.63926 & 16.49  & 3.43 & Planet Finder Spectrograph\\
2460154.60562 & 5.10   & 2.63& Planet Finder Spectrograph \\
2460154.61980 & 8.69   & 2.45 & Planet Finder Spectrograph\\
2460154.63393 & 4.43   & 2.66& Planet Finder Spectrograph \\
2460155.64044 & 5.92   & 2.36 & Planet Finder Spectrograph\\
2460155.65419 & 2.74   & 2.21& Planet Finder Spectrograph \\
2460155.66844 & 14.53  & 2.18 & Planet Finder Spectrograph\\
2460156.65824 & -8.11  & 2.47 & Planet Finder Spectrograph\\
2460156.67302 & -7.01  & 2.44& Planet Finder Spectrograph \\
2460156.68696 & 9.74   & 2.66 & Planet Finder Spectrograph\\
2460160.66059 & -33.60 & 2.68& Planet Finder Spectrograph \\
2460160.67464 & -38.21 & 2.89& Planet Finder Spectrograph \\
2460160.68876 & -31.25 & 2.81& Planet Finder Spectrograph \\
2460180.53256 & -14.42 & 3.48& Planet Finder Spectrograph \\
2460180.54735 & 2.14   & 3.30& Planet Finder Spectrograph \\
2460180.56107 & -16.94 & 3.67 & Planet Finder Spectrograph\\
2460184.54002 & -40.53 & 4.71& Planet Finder Spectrograph \\
2460210.52069 & -7.26  & 3.66& Planet Finder Spectrograph \\
2460210.53530 & -5.42  & 3.68& Planet Finder Spectrograph \\
2460210.54950 & -8.28  & 3.58& Planet Finder Spectrograph \\
2460212.51455 & 2.93   & 3.17 & Planet Finder Spectrograph\\
2460212.52869 & 0.00   & 2.76 & Planet Finder Spectrograph\\
2460212.54273 & -5.89  & 2.62& Planet Finder Spectrograph\\
2459819.56147 & 5965.0 & 32.0 & CHIRON \\
2459822.57141 & 5992.0 & 28.0 & CHIRON \\
2459825.47184 & 6030.0 & 28.0 & CHIRON \\
2459828.63916 & 5984.0 & 31.0 & CHIRON \\
2459838.46341 & 5994.0 & 26.0 & CHIRON \\
2459841.54330 & 5984.0 & 31.0 & CHIRON\\
2459844.54869 & 6038.0 & 32.0 & CHIRON \\
2459854.51278 & 6047.0 & 32.0 & CHIRON\\
2459857.51441 & 6015.0 & 31.0 & CHIRON\\
2459860.49822 & 5980.0 & 38.0 & CHIRON\\
2459863.50449 & 5997.0 & 35.0 & CHIRON\\
2459866.52502 & 6006.0 & 28.0 & CHIRON\\
2459869.50415 & 6017.0 & 33.0 & CHIRON\\
2459872.52953 & 6033.0 & 38.0 & CHIRON\\
2459876.50716 & 6070.0 & 48.0 & CHIRON \\
2459680.91764 & 5393.1 & 35.7 & CORALIE  \\
2459690.91570 & 5436.0 & 61.8 & CORALIE  \\
2459702.81753 & 5354.2 & 37.8& CORALIE \\
2459723.87356 & 5428.1 & 54.7& CORALIE \\
2459740.79469 & 5340.0 & 70.5& CORALIE \\
2459748.77041 & 5254.7 & 93.9& CORALIE \\
2459845.62784 & 5329.9 & 49.8 & CORALIE\\
\enddata
\end{deluxetable*}

\begin{deluxetable*}{lccc}
\tablenum{4} \tablecolumns{4}
\caption{Radial Velocity Datasets (Continued)}
\label{tab:RVs}
\tablehead{
\colhead{Time (BJD)} & \colhead{RV (m/s)} & \colhead{RV error (m/s)} & \colhead{Instrument}}
\startdata
2459763.78280 & 5349.6 & 14.6 & FEROS \\
2459764.78507 & 5330.3 & 11.4 & FEROS \\
2459806.64321 & 5340.8 & 20.8 & FEROS\\
2459808.62570 & 5358.8 & 10.1& FEROS \\
2459809.63296 & 5319.1 & 13.4 & FEROS\\
2459810.67493 & 5291.7 & 12.0 & FEROS\\
2459835.57461 & 5413.7 & 14.6 & FEROS\\
2459847.63789 & 5393.9 & 17.4 & FEROS\\
2459845.65994 & 5368.4 & 12.9 & FEROS \\
2459851.60212 & 5363.3 & 14.1 & FEROS\\
2459897.50622 & 5303.8 & 11.9 & FEROS\\
2460032.88150 & 5288.0 & 10.3 & FEROS\\
2460047.87740 & 5376.4 & 9.3 & FEROS \\
2460066.86333 & 5401.1 & 9.8  & FEROS\\
2460125.74290 & 5320.0 & 8.7  & FEROS\\
2460149.76750 & 5331.6 & 9.9  & FEROS\\
2460151.83118 & 5352.2 & 10.7 & FEROS\\
2460178.52971 & 5335.2 & 12.1 & FEROS\\
2460193.51918 & 5323.4 & 11.4 & FEROS\\
2460194.69004 & 5445.8 & 16.0 & FEROS\\
2460227.54377 & 5371.7 & 4.3	& HARPS-1\\    
2460239.52640 & 5402.2 &  4.8	& HARPS-1\\   
2460254.54061 & 5392.2 & 4.1	& HARPS-1\\    
2460262.53137 & 5391.3 & 3.7	& HARPS-1\\   
2460056.82106 & 5233.7 & 6.4 & HARPS-2\\  
2460058.85810 & 5242.8 & 8.2 & HARPS-2\\  
2460060.87608 & 5232.6 &7.9& HARPS-2\\
2460062.90944 & 5237.2 & 7.4	& HARPS-2\\
2460086.79806 & 5234.7 & 11.6 & HARPS-2\\
2460089.81426 & 5250.6 & 6.3 & HARPS-2\\
\enddata
\end{deluxetable*}

\section{Radial Velocity Observations}\label{sec:appendix}

\bibliography{references}{}
\bibliographystyle{aasjournal}

\end{document}